\documentclass[%
reprint,
superscriptaddress,
 amsmath,amssymb,
 aps, physrev,
]{revtex4-2}
\usepackage{extarrows}
\usepackage{color}
\usepackage{xcolor}
\usepackage{tabularx}
\usepackage{array}
\usepackage{multirow}
\newcolumntype{Y}{>{\centering\arraybackslash}X}
\usepackage{graphicx}% Include figure files
\usepackage{dcolumn}% Align table columns on decimal point
\usepackage{siunitx}
\usepackage{bm}% bold math
\usepackage[version=4]{mhchem}
\usepackage{hyperref}% add hypertext capabilities
\hypersetup{colorlinks=true,linkcolor=blue,citecolor=blue,urlcolor=blue}
\usepackage{cleveref}
\usepackage[colorinlistoftodos]{todonotes}
\usepackage{soul} %\st delete, \hl, highlight, \ul, _, \caps

\begin{document}
\title{
Controllable charge-spin conversion in sliding multiferroic metal \ce{Fe3GaTe2}
}

\author{Tongrui Li}
\affiliation{Hefei National Research Center for Physical Sciences at Microscale and School of Emerging Technology, University of Science and Technology of China, Hefei, 230026, China}
\affiliation{National Synchrotron Radiation Laboratory, University of Science and Technology of China, Hefei, Anhui 230029, China}

\author{Lidong Zhang} 
\affiliation{National Synchrotron Radiation Laboratory, University of Science and Technology of China, Hefei, Anhui 230029, China}

\author{Weiyi Pan}
\email{weiyi.pan@physik.uni-regensburg.de}
\affiliation{Institute for Theoretical Physics, University of Regensburg, 93040 Regensburg, Germany}

\begin{abstract} 
Controlling charge--spin conversion within a single magnetic material is important for compact spintronic devices. Using first-principles calculations, symmetry analysis, and Wannier-based linear response, we investigate the Edelstein and spin Hall responses of sliding-ferroelectric bilayer \ce{Fe3GaTe2}. We identify two degenerate ground states with opposite interlayer sliding vectors and oppositely directed out-of-plane dipole moments, separated by a calculated switching barrier of approximately $16.3$~meV per unit cell. Magnetization orientation determines the allowed current-induced spin components, whereas ferroelectric switching produces component-selective sign reversals governed by $M_z\mathcal T$ for in-plane magnetization and by $M_z$ for out-of-plane magnetization. The Edelstein response is strongly energy dependent, with the induced spin accumulation reaching nearly $2~\hbar$ for $\mathbf m\parallel+y$ under the reference field of $1$~V/\AA. Momentum-resolved analysis associates the time-reversal (TR) even and TR-odd Edelstein responses with $\Gamma$-centered constant-energy contours and localized interband features, respectively. The spin-conductivity contributions exhibit the opposite momentum-space character: broad, sign-competing distributions of spin Berry curvature for TR-even and narrow Fermi-surface contours for TR-odd. The transverse spin Hall conductivity reaches values of order $10^{3}\,(\hbar/e)(\mathrm{S/cm})$ and depends strongly on both the ferroelectric state and magnetization direction. Continuous magnetization rotations in the $xy$ and $xz$ planes further reveal symmetry-allowed higher angular harmonics in both spin accumulation and spin-current responses. These results establish sliding-ferroelectric bilayer \ce{Fe3GaTe2} as an effective platform for tunable charge--spin conversion. The combined ferroelectric and magnetic tunability extends the functionality of metallic multiferroics to nonequilibrium spin control and offers a practical materials strategy for reconfiguring both spin accumulation and spin-current generation.
\end{abstract}

\maketitle
\clearpage
%---------------------------introduction-------------------------
\section{Introduction}

Charge--spin conversion in spintronics is commonly realized through two complementary responses: the Edelstein effect (EE) and the spin Hall effect (SHE). In the EE, an electric field induces a nonequilibrium spin accumulation in a spin--orbit-coupled (SOC), inversion-asymmetric system~\cite{Edelstein1990,Johansson2024EdelsteinReview,xly2mhk9}, whereas the SHE converts a longitudinal charge current into a transverse spin current~\cite{Hirsch1999SHE,Sinova2015SHE,Guo2008PtSHE}. Although both responses are governed by SOC and crystal symmetry, they produce different observables: a local spin density in the EE and spin-current transport in the SHE~\cite{Sayed2021Interconversion,Han2018QuantumMaterials}. Interface engineering, electrostatic gating, and twist angle provide established routes for realizing and tuning spin--charge interconversion~\cite{Soumyanarayanan2016SOC,Benitez2020TunableSpin,Lee2022Twisted,Vaz2019Oxide2DEG,Varotto2022KTaO3,Lesne2016Oxide}. In nonmagnetic materials, crystallographic symmetry determines which EE and SHE tensor components are allowed~\cite{Johansson2024EdelsteinReview,Seemann2015ResponseTensors,Pan2026PtBi2}. In magnetic materials, the magnetization direction and magnetic structure can modify the magnetic point group and thereby selectively activate or suppress response components and tune their magnitudes, while exchange fields can further enhance the EE~\cite{Lazrak2024ExchangeEdelstein,Pan2026VTe2,Qu2020MagnetizationSHE,Chen2021AFMSHE,Zhu2026MagneticStructureSHE}. The EE and spin-conductivity tensors can additionally be decomposed into components that are even and odd under reversal of the magnetic order~\cite{Pan2026VTe2,Zheng2024AnatomySHE}. The resulting spin accumulation and spin currents can generate spin--orbit torques that enable electrical switching of magnetic order and provide a basis for current-driven nonvolatile memory and logic~\cite{Miron2011SOT,Liu2012TaSOT,Zelezny2017SOTSymmetry,Manchon2019SOT}.

Beyond magnetic control, switchable polar order provides an effective and potentially nonvolatile knob for charge--spin conversion. In \ce{GeTe}, ferroelectric polarization controls the Rashba splitting and spin texture~\cite{DiSante2013GeTe,Liebmann2016GeTe,Rinaldi2018GeTe}, and enables room-temperature nonvolatile switching of spin-to-charge conversion~\cite{Varotto2021GeTe}. Nonvolatile reversal of spin-to-charge conversion has also been demonstrated in an electric-field-induced ferroelectric-like \ce{SrTiO3}-based Rashba system~\cite{Noel2020Nonvolatile}. First-principles studies further predict polarization-controlled EE in \ce{WTe2} and monolayer \ce{PtBi2}~\cite{Jafari2022WTe2,Pan2026PtBi2}. The linear intrinsic spin Hall conductivity is invariant under spatial inversion because the spin-current operator and electric field are both inversion odd; two domains related solely by inversion therefore have identical spin Hall conductivity tensors~\cite{Seemann2015ResponseTensors,Pan2026PtBi2Bilayer}. Nonvolatile control of the spin Hall response remains possible when switching changes the polarization configuration or interfacial electronic structure, as demonstrated for Pt/PZT interfaces and predicted for metal--ferroelectric multilayers and bilayer \ce{PtBi2}~\cite{Fang2020FerroelectricSOC,Xu2022FerroelectricSHC,Pan2026PtBi2Bilayer}. Changes in the polar-distortion amplitude or structural phase can likewise modify the SHC of \ce{GeTe} and \ce{SnTe}~\cite{Wang2020GeTeSnTeSHE,Zhang2020GeTeSHC}.

A conducting material with switchable polarization and magnetic order would place both control knobs in a single system, enabling versatile reconfiguration of charge--spin conversion. Conventional multiferroics, however, are usually insulating and therefore cannot directly serve as current-carrying active layers for the EE or SHE; their scarcity also reflects the competing electronic requirements for conventional ferroelectric distortions and magnetic transition-metal ions~\cite{Hill2000MagneticFerroelectrics,Eerenstein2006Multiferroic,Cheong2007Multiferroics,Fiebig2016MultiferroicsEvolution,Spaldin2019Multiferroics}. Carrier doping can introduce conductivity, but charged dopants and defects can perturb polar distortions in a strongly material-dependent manner~\cite{Chandra2013DopedFerroelectrics,Xia2019DopedFerroelectrics}. Although mobile carriers screen long-range electrostatic fields, polar metals show that metallicity does not preclude stable inversion breaking; electrical switchability nevertheless remains uncommon~\cite{Anderson1965FerroelectricMetals,Shi2013LiOsO3,Benedek2016FerroelectricMetals,Bhowal2023PolarMetals}. Sliding ferroelectricity offers an alternative route to switchable out-of-plane polarization in van der Waals bilayers: an in-plane translation changes the stacking registry and reverses the polarization through interlayer charge redistribution~\cite{Li2017VerticalPolarization,Wu2021SlidingReview,Ji2023StackingTheory,Wang2023VdWFerroelectrics}. It has been realized in hBN and transition-metal dichalcogenide homo- and heterobilayers~\cite{Yasuda2021BN,ViznerStern2021Sliding,Wang2022TMD,Rogee2022Heterobilayers}. Ferroelectric switching has also been demonstrated in conducting few-layer \ce{WTe2}~\cite{Fei2018WTe2}, while metallic sliding ferroelectricity has been predicted in $M\mathrm{Te}_2$ bilayers ($M=\mathrm{Pt},\mathrm{Pd},\mathrm{Ni}$)~\cite{Sheng2023FerroelectricMetals}. When combined with two-dimensional magnetism, this mechanism can yield sliding multiferroics, including metallic ones such as bilayer \ce{Fe3GeTe2}~\cite{Zhong2023MagneticSliding,Miao2024SlidingFGT}, suggesting an intrinsic route to nonvolatile, multichannel control of charge--spin conversion. Within this design space, \ce{Fe3GaTe2} is especially attractive. It is a metallic van der Waals ferromagnet with strong perpendicular magnetic anisotropy and a Curie temperature above room temperature~\cite{Zhang2022Fe3GaTe2,Ruiz2024Fe3GaTe2Origin,Wu2024WaferFe3GaTe2,Marfoua2024Fe3GaTe2Transport}, and current-driven spin--orbit-torque switching has been demonstrated in \ce{Fe3GaTe2}-based heterostructures~\cite{Li2023Fe3GaTe2SOT,Kajale2024AllVdWSOT}. Its low-energy bands are dominated by spin-polarized Fe-$3d$ states, and recent calculations for monolayer \ce{Fe3GaTe2} reveal strongly sublattice- and momentum-dependent current-induced torques~\cite{Brizolla2026SOT}. Whether sliding polarization and magnetization orientation can jointly and selectively control both the EE and SHE in an intrinsic \ce{Fe3GaTe2} bilayer remains unresolved.

Here, using first-principles calculations, symmetry analysis, and Wannier-based Kubo linear response, we investigate charge--spin conversion in sliding-ferroelectric bilayer \ce{Fe3GaTe2}. We identify two degenerate ground states connected by opposite interlayer sliding and determine how ferroelectric switching modifies the Edelstein and spin-conductivity responses. The magnetization dependence is examined along high-symmetry directions and during continuous rotations in the $xy$ and $xz$ planes. Momentum-resolved calculations further reveal the electronic origins and symmetry constraints of the induced spin accumulation and spin-current responses. Our results establish sliding-ferroelectric bilayer \ce{Fe3GaTe2} as an effective multiferroic-metal platform for tunable charge--spin conversion and identify ferroelectric switching and magnetization rotation as complementary means of reconfiguring spin accumulation and spin-current generation within a single conducting magnetic material.

\section{Computational Methods}

The electronic structure of bilayer \ce{Fe3GaTe2} was obtained from density functional theory calculations using the projector augmented-wave method~\cite{Blochl1994PAW} implemented in VASP~\cite{Kresse1996VASP}. The exchange--correlation energy was described within the local-density approximation (LDA)~\cite{Perdew1981LDA}. The in-plane lattice parameters were fixed at $a=b=3.948$~\AA, as obtained from the optimized monolayer \ce{Fe3GaTe2}, while the atomic coordinates of the bilayer were allowed to relax. The structural optimization was carried out in the ferromagnetic state without SOC. A plane-wave cutoff energy of 500~eV and a $16\times16\times1$ $\mathbf{k}$-point mesh were used. For each ionic step, the electronic self-consistent cycle was converged until the total-energy change between successive electronic steps was below $10^{-7}$~eV. The atomic positions were then relaxed until the Hellmann--Feynman forces on all atoms were below $10^{-3}$~eV/\AA.

To evaluate the switching barrier between the two ferroelectric states, the minimum-energy path connecting the two opposite sliding configurations was determined using the climbing-image nudged elastic band (CI-NEB) method~\cite{Henkelman2000CINEB}. The two relaxed ferroelectric structures were taken as the initial and final states, and eight intermediate images were inserted between them, giving ten images along the complete switching path. During the CI-NEB calculation, the atomic coordinates of the intermediate images were relaxed along the minimum-energy path, while the climbing image was driven toward the saddle point. The switching barrier was obtained from the energy difference between the saddle point and the ferroelectric ground state.

Starting from the relaxed bilayer structure, separate self-consistent calculations including SOC were performed for each magnetization orientation. Since bilayer \ce{Fe3GaTe2} is metallic, the Brillouin zone was sampled using a denser $22\times22\times1$ $\mathbf{k}$ mesh in these calculations. The magnetization direction was specified independently for each configuration, including the high-symmetry directions and the continuous rotations considered in the $xy$ and $xz$ planes. The magnetization-dependent SOC calculations were performed following Ref.~\cite{Pan2026VTe2}.

For each magnetic configuration, the SOC Bloch states were transformed to a Wannier representation using Wannier90 \cite{Mostofi2008Wannier90}. Fe-$d$, Ga-$p$, and Te-$p$ orbitals were chosen as the projection basis. The bilayer unit cell therefore contains 48 spatial orbitals, corresponding to 96 spinor Wannier functions. The resulting tight-binding Hamiltonian was subsequently symmetrized using WannSymm~\cite{Zhi2022WannSymm}. The symmetrization was performed according to the magnetic symmetry associated with the corresponding magnetization orientation, thereby suppressing small numerical violations of the symmetry relations in the interpolated Hamiltonian.

The current-induced spin accumulation was calculated from the symmetrized Wannier Hamiltonian using the LinRes code \cite{ZeleznyLinres}. Within linear response, the induced spin
polarization is written as
\begin{equation}
\delta S_i=\sum_j \chi_{ij}E_j ,
\label{eq:EE_def}
\end{equation}
where $\chi_{ij}$ denotes the Edelstein susceptibility. For a magnetic
system, the response can be separated according to its behavior under
magnetization reversal,
\begin{equation}
\chi_{ij}
=
\chi_{ij}^{\mathrm{even}}
+
\chi_{ij}^{\mathrm{odd}},
\label{eq:EE_decomposition}
\end{equation}
with
$\chi_{ij}^{\mathrm{even}}(-\mathbf m)
=\chi_{ij}^{\mathrm{even}}(\mathbf m)$ and
$\chi_{ij}^{\mathrm{odd}}(-\mathbf m)
=-\chi_{ij}^{\mathrm{odd}}(\mathbf m)$.
Within the Kubo linear-response formalism
\cite{Freimuth2014SOT,Pan2026PtBi2,Pan2026VTe2}, these two
contributions are evaluated as
\begin{equation}
\chi_{ij}^{\mathrm{even}}
=
-\frac{e~\hbar}{\pi N_k}
\sum_{\mathbf{k},m,n}
\frac{
\Gamma^2
\,{\rm Re}
\left[
\langle n\mathbf{k}|\hat S_i|m\mathbf{k}\rangle
\langle m\mathbf{k}|\hat v_j|n\mathbf{k}\rangle
\right]
}{
\left[(E_F-\varepsilon_{n\mathbf{k}})^2+\Gamma^2\right]
\left[(E_F-\varepsilon_{m\mathbf{k}})^2+\Gamma^2\right]
},
\label{eq:EE_even}
\end{equation}
and
\begin{equation}
\begin{aligned}
\chi_{ij}^{\mathrm{odd}}
={}&
\frac{2e~\hbar}{N_k}
\sum_{\mathbf{k}}
\sum_{n\in{\rm occ.}}
\sum_{\substack{m\in{\rm unocc.}\\m\neq n}}
{\rm Im}
\left[
\langle n\mathbf{k}|\hat S_i|m\mathbf{k}\rangle
\langle m\mathbf{k}|\hat v_j|n\mathbf{k}\rangle
\right]
\\
&\times
\frac{
\Gamma^2-
(\varepsilon_{n\mathbf{k}}-\varepsilon_{m\mathbf{k}})^2
}{
\left[
(\varepsilon_{n\mathbf{k}}-\varepsilon_{m\mathbf{k}})^2
+\Gamma^2
\right]^2}.
\end{aligned}
\label{eq:EE_odd}
\end{equation}
Here, $N_k$ is the number of sampled $\mathbf{k}$ points, $\varepsilon_{n\mathbf{k}}$ is the band energy, $\hat S_i$ and $\hat v_j$ are the spin and velocity operators, respectively, and $\Gamma$ is the phenomenological broadening. We use $\Gamma=10$ meV, which corresponds to $\tau=~\hbar/(2\Gamma)\simeq33$ fs within this constant-broadening convention and is comparable to the values adopted in recent Wannier-based calculations of current-induced spin responses~\cite{Freimuth2014SOT,Pan2026VTe2,Pan2026PtBi2}. More details are provided in Appendix K. A $600\times600\times1$ $\mathbf{k}$ mesh was employed for the linear-response integration.

Throughout this work, the electric field is taken along the $+x$
direction. Hence only the first column of the Edelstein tensor is
required,
\begin{equation}
\delta S_i^{\mathrm{even/odd}}
=
\chi_{ix}^{\mathrm{even/odd}}E_x .
\label{eq:deltaS_Ex}
\end{equation}
We set a reference field of
$E_x=10^{10}$~V/m $=1$~V/\AA.
Accordingly, the results are presented directly as the induced spin accumulations $\delta S_i$, rather than as the susceptibility coefficients $\chi_{ix}$. The reference field is used only to define a common response scale and does not imply the experimentally applied field strength. A similar convention has been used in previous calculations of current-induced spin accumulation in two-dimensional magnetic systems~\cite{Pan2026VTe2}.

The spin-conductivity response was evaluated from the same Wannier Hamiltonians. We define
\begin{equation}
J_i^{\alpha}
=
\sigma_{ij}^{\alpha}E_j ,
\label{eq:SHE_def}
\end{equation}
where $J_i^\alpha$ denotes a spin current propagating along direction
$i$ with spin polarization along $\alpha$. The spin conductivity was
decomposed into contributions that are even and odd under magnetization
reversal. Within the same constant-broadening Kubo formalism
\cite{Pan2026VTe2,Freimuth2014SOT}, these contributions are

\begin{equation}
\begin{aligned}
\sigma_{ij}^{\alpha,\mathrm{odd}}
={}&
-\frac{e~\hbar}{\pi N_k V_{\mathrm{eff}}}
\sum_{\mathbf{k},m,n}
\Gamma^2
\operatorname{Re}
\left[
\langle n\mathbf{k}|\hat{J}_i^\alpha|m\mathbf{k}\rangle
\langle m\mathbf{k}|\hat{v}_j|n\mathbf{k}\rangle
\right]
\\
&\times
\frac{1}{
\left[
(\mu-\varepsilon_{n\mathbf{k}})^2+\Gamma^2
\right]
\left[
(\mu-\varepsilon_{m\mathbf{k}})^2+\Gamma^2
\right]
}.
\end{aligned}
\label{eq:spin_conductivity_odd}
\end{equation}

\begin{equation}
\begin{aligned}
\sigma_{ij}^{\alpha,\mathrm{even}}
={}&
\frac{2e~\hbar}{N_k V_{\mathrm{eff}}}
\sum_{\mathbf{k}}
\sum_{n\in\mathrm{occ.}}
\sum_{m\in\mathrm{unocc.}}
\\
&\operatorname{Im}
\left[
\langle n\mathbf{k}|\hat{J}_i^\alpha|m\mathbf{k}\rangle
\langle m\mathbf{k}|\hat{v}_j|n\mathbf{k}\rangle
\right]
\\
&\times
\frac{
\Gamma^2-
(\varepsilon_{n\mathbf{k}}-\varepsilon_{m\mathbf{k}})^2
}{
\left[
(\varepsilon_{n\mathbf{k}}-\varepsilon_{m\mathbf{k}})^2
+\Gamma^2
\right]^2
}.
\end{aligned}
\label{eq:spin_conductivity_even}
\end{equation}

The spin-current operator is defined as

\begin{equation}
\hat{J}_i^\alpha
=
\frac{1}{2}
\left\{
\hat{v}_i,\hat{S}_\alpha
\right\}.
\label{eq:spin_current_operator}
\end{equation}

The time-reversal-even contribution is interband dominated and
approaches the intrinsic spin-Berry-curvature expression in the
small-broadening limit. The time-reversal-odd contribution is
dominated by intraband Fermi-surface transport and exhibits the
corresponding broadening dependence. For the reported spin conductivities, $V_{\mathrm{eff}}=A d_{\mathrm{eff}}$ is the effective normalization volume, where $A$ is the in-plane unit-cell area and $d_{\mathrm{eff}}=12.96~\text{\AA}$ is the effective bilayer thickness defined in Appendix~D. Equivalently, the conductivities initially normalized by the periodic supercell volume $Ac$ were rescaled by the factor $c/d_{\mathrm{eff}}$. For an electric field along $+x$, the calculations include both the longitudinal components $\sigma_{xx}^{\alpha}$ and the transverse components $\sigma_{yx}^{\alpha}$. The complete set is referred to as the spin-conductivity tensor. All spin-conductivity calculations were performed on a $600\times600\times1$ $\mathbf{k}$-point mesh.

\begin{figure}[t]
\includegraphics[width=\columnwidth]{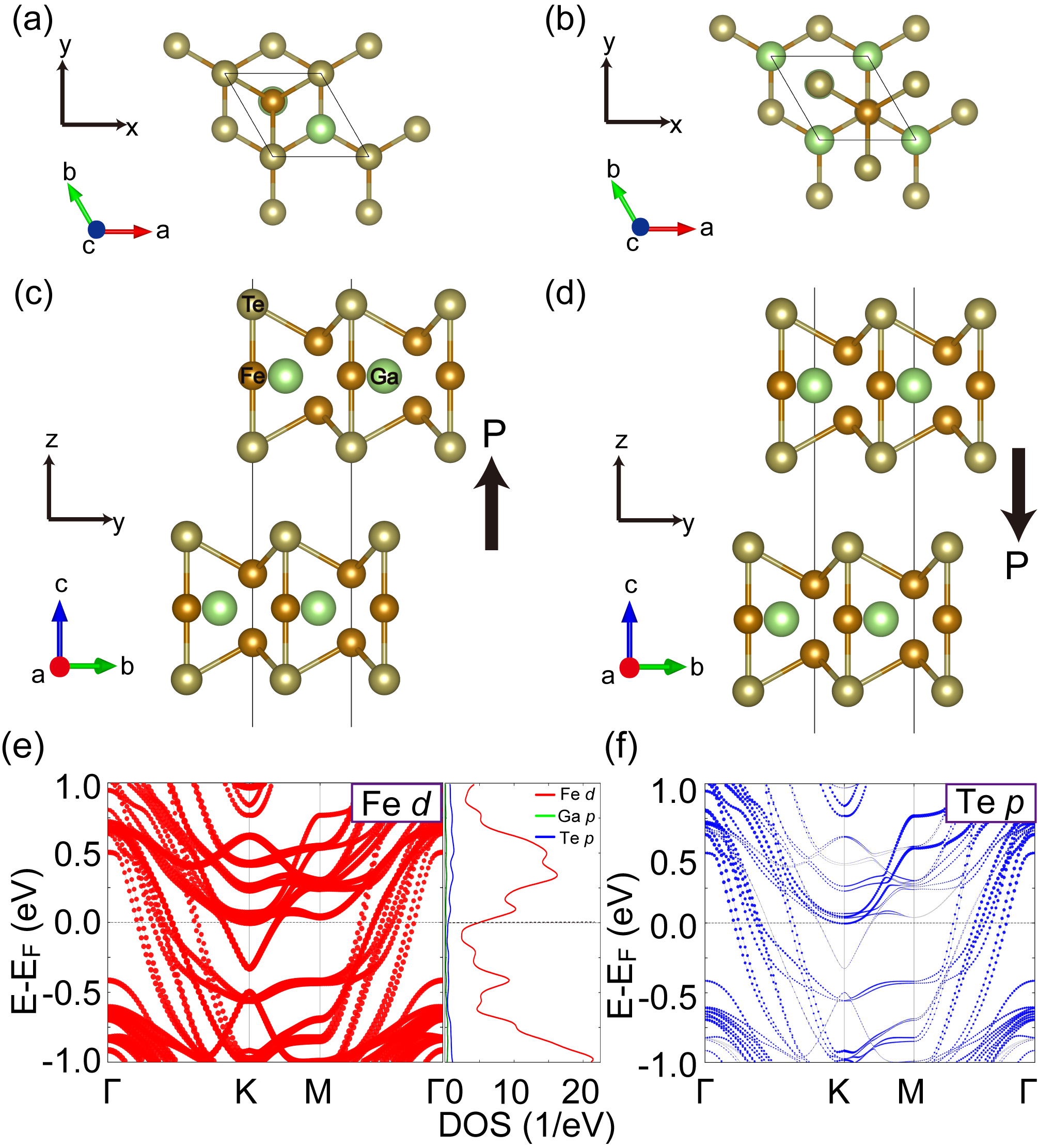}
\caption{ Crystal structure and electronic structure of bilayer \ce{Fe3GaTe2}. (a,b) Top views of the two ferroelectric configurations with opposite interlayer sliding. (c,d) Corresponding side views of the two configurations. The black arrows indicate the directions of the out-of-plane electric polarization, which are reversed between the two states. (e,f) Orbital-projected electronic band structures along the $\Gamma$--K--M--$\Gamma$ path, highlighting the Fe-$d$ and Te-$p$ contributions, respectively. The size of the colored markers represents the corresponding orbital weight. The projected density of states is shown on the right side of (e), with Fe-$d$, Ga-$p$, and Te-$p$ contributions shown in red, green, and blue, respectively. }
\label{fig1}
\end{figure}

\section{Results and Discussion}

\subsection{Sliding ferroelectric structure and electronic states}

Figures~\ref{fig1}(a)--\ref{fig1}(d) show the crystal structures of bilayer \ce{Fe3GaTe2}. The bilayer belongs to the space group $P3m1$ (No.~156) with $C_{3v}$ point-group symmetry. Two degenerate polarization states are obtained by reversing the relative in-plane displacement between the two layers. We denote the state with a positive out-of-plane dipole moment as $P_{\uparrow}$ and that with a negative dipole moment as $P_{\downarrow}$. As illustrated by the side views in Figs.~\ref{fig1}(c) and \ref{fig1}(d), the two states carry opposite out-of-plane electric dipole moments. The corresponding sliding vectors are $\mathbf{t}_{P_{\uparrow}}=(1/3,-1/3)$ and $\mathbf{t}_{P_{\downarrow}}=(-1/3,1/3)$, with $p_z=+2.19\times10^{-4}\,e\AA$ and $p_z=-2.19\times10^{-4}\,e\AA$, respectively. As summarized in Table~\ref{table1}, the $P_{\uparrow}$ and $P_{\downarrow}$ states are energetically degenerate and constitute the two lowest-energy sliding configurations, whereas the other high-symmetry sliding configurations lie approximately $46$~meV per unit cell higher in energy. The magnetic anisotropy favors an out-of-plane magnetization, with $\Delta E_{\mathrm{MAE}} =E_{\mathrm{in\text{-}plane}}-E_{\mathrm{out\text{-}of\text{-}plane}} \simeq 4.08$~meV per unit cell ($0.68$~meV/Fe), indicating an out-of-plane magnetic easy axis. The two polarization states can be switched through relative interlayer sliding. Our CI-NEB calculation gives an energy barrier of approximately $16.3$ meV per unit cell along the minimum-energy path, as shown in Fig.~\ref{fig14} and discussed in Appendix I. The existence of a low-energy sliding path is consistent with theoretical studies of sliding-ferroelectric van der Waals bilayers, including magnetic bilayer Fe$_3$GeTe$_2$~\cite{Li2017VerticalPolarization,Ji2023StackingTheory,Zhong2023MagneticSliding,Miao2024SlidingFGT}. Appendix J shows that ferromagnetic interlayer coupling remains favorable over the relevant range of van der Waals gaps. The calculated barrier does not by itself determine an experimental
coercive field. Ferroelectric reversal generally proceeds through
spatially nonuniform processes, including local nucleation and
domain-wall propagation~\cite{He2023CIPS,Shin2007DomainWall}. A more detailed discussion is provided in Appendix I.

Figures~\ref{fig1}(e) and \ref{fig1}(f) show the orbital-projected electronic structures of the $P_{\uparrow}$ ferroelectric state. Consistent with reported electronic structures of monolayer Fe$_3$GaTe$_2$, the states around the Fermi level are dominated by spin-polarized Fe-$3d$ orbitals, while Te-$p$ states provide a smaller contribution~\cite{Zhang2022Fe3GaTe2,Ruiz2024Fe3GaTe2Origin,Brizolla2026SOT}. The ferromagnetic state produces a pronounced exchange-split multiband structure near the Fermi level, with several Fe-$d$-derived bands crossing or approaching $E_F$. In the bilayer, the additional sliding-induced polarization asymmetry and interlayer electronic reconstruction provide further ingredients that can modify the current-induced spin polarization. It is therefore not evident from the band structure alone whether the Edelstein response retains the behavior found in the monolayer or develops qualitatively different features in the polarization bilayer. To address this point, we first examine the energy-dependent Edelstein spin accumulation for different magnetization directions.

\begin{table}[htbp]
\centering
\caption{
Relative energies of bilayer \ce{Fe3GaTe2} for different interlayer
sliding configurations. The sliding vector denotes the in-plane
translation of the upper layer relative to the lower layer.
States 01 and 02 correspond to the degenerate
$P_{\uparrow}$ and $P_{\downarrow}$ states, respectively,
with opposite sliding vectors and out-of-plane dipole moments.
The lowest energy is taken as the reference.
}
\label{table1}
\begin{tabular}{ccc}
\hline\hline
State & Sliding operation & $\Delta E$ (meV) \\
\hline
01 & $(1/3,-1/3)$  & 0.00  \\
02 & $(-1/3,1/3)$  & 0.00  \\
03 & $(1/3,0)$      & 46.41 \\
04 & $(-1/3,0)$     & 46.41 \\
05 & $(0,1/3)$      & 46.41 \\
06 & $(0,-1/3)$     & 46.41 \\
07 & $(1/3,1/3)$    & 46.41 \\
08 & $(-1/3,-1/3)$  & 46.20 \\
\hline\hline
\end{tabular}
\end{table}

\begin{figure}[t]
\includegraphics[width=\columnwidth]{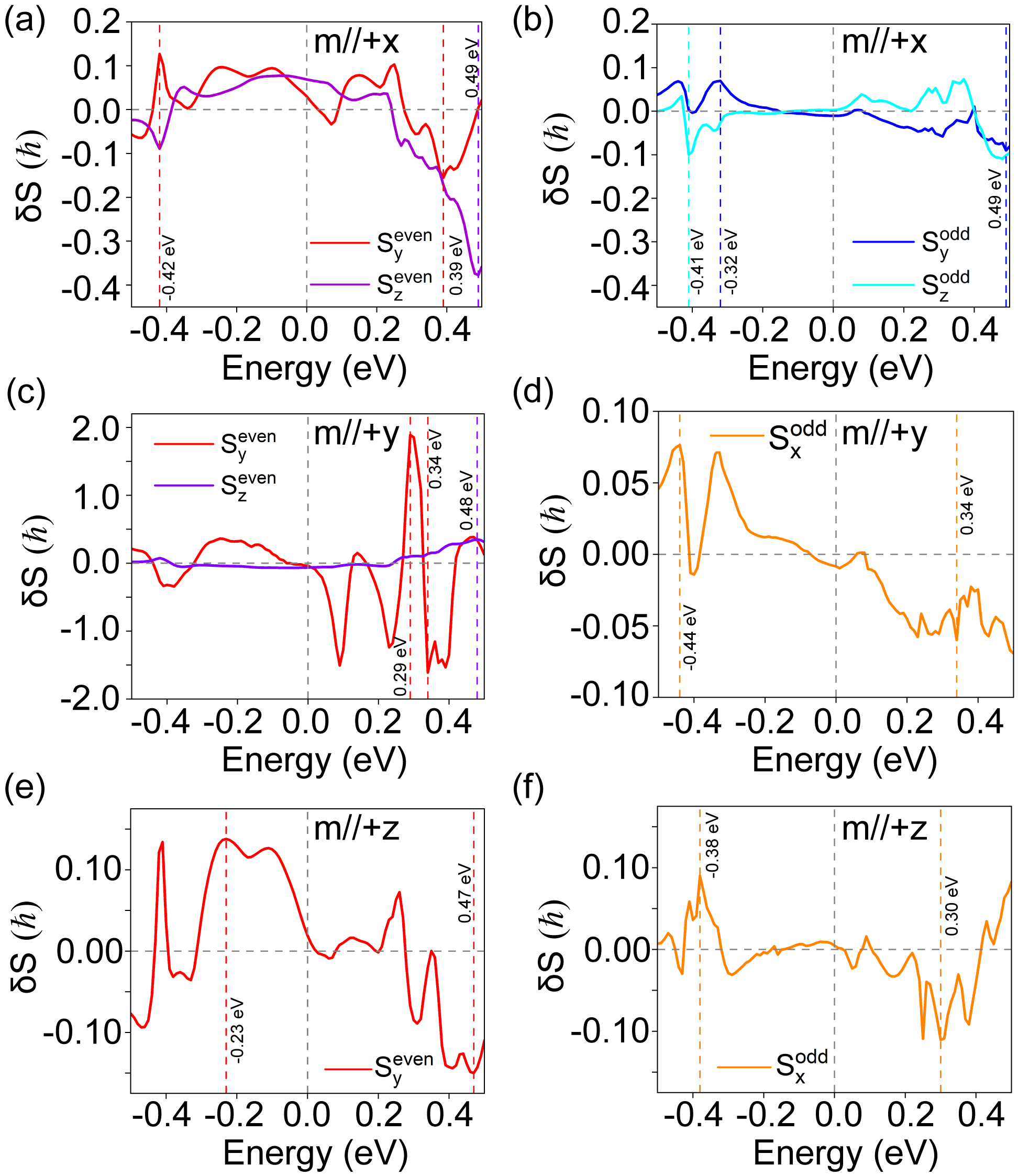}
\caption{ Energy dependence of the total current-induced spin accumulation in $P_{\uparrow}$ state under an electric field applied along the $+x$ direction, with the reference field $E_x=1$~V/\AA. (a,b) Time-reversal-even and -odd components, respectively, for $\mathbf m\parallel +x$. (c,d) Corresponding even and odd components for $\mathbf m\parallel +y$. (e,f) Corresponding even and odd components for $\mathbf m\parallel +z$. The solid curves show the total spin accumulation summed over all atomic contributions. The colored vertical dashed lines indicate selected representative energies used for the momentum-resolved analysis, with their colors corresponding to the associated spin components. }
\label{fig2}
\end{figure}

\begin{figure*}[t]
\includegraphics[width=170mm]{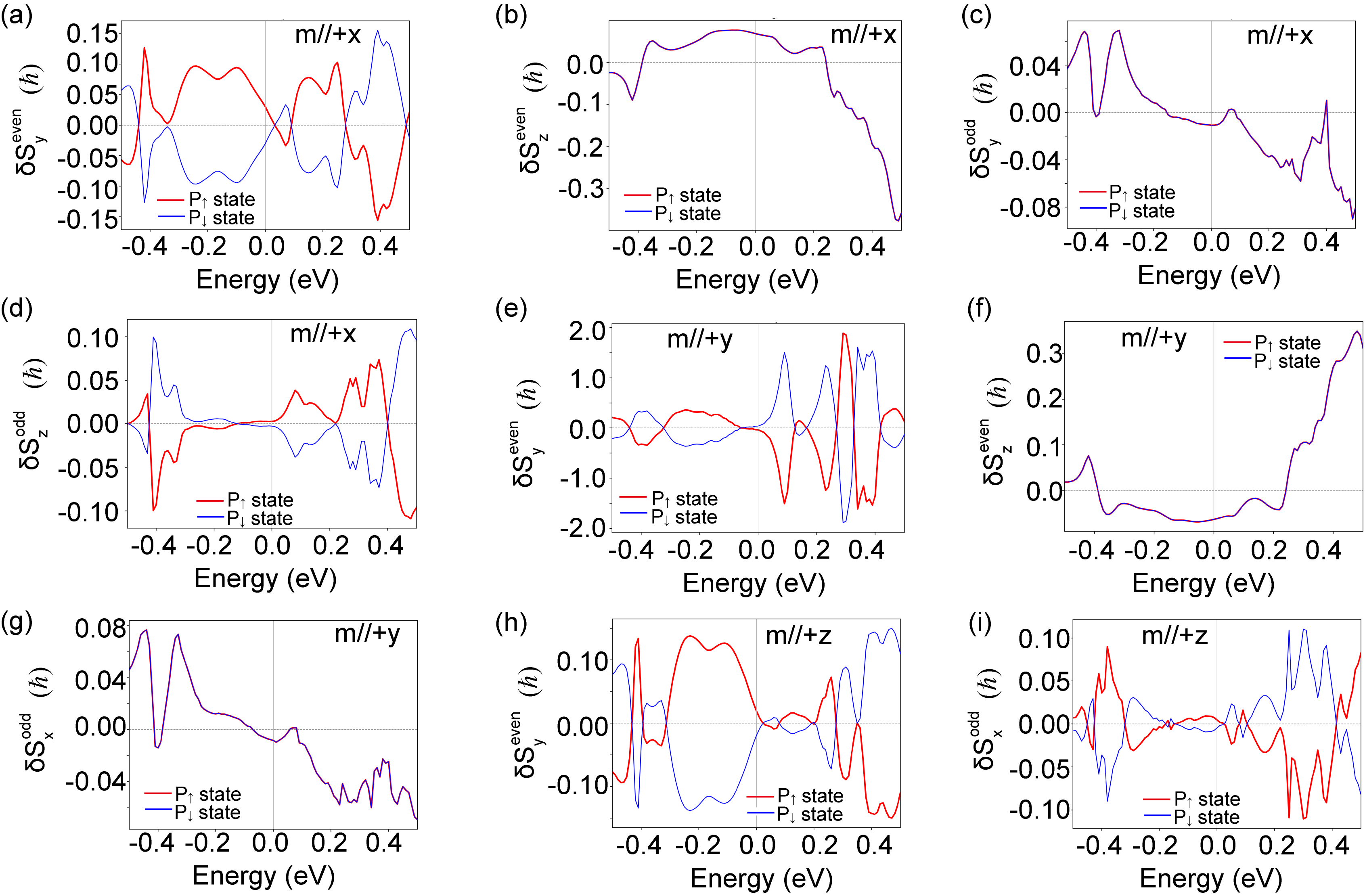}
\caption{ Ferroelectric-state dependence of the current-induced spin accumulation under an electric field applied along the $+x$ direction, with the reference field $E_x=1$~V/\AA. Red and blue curves denote $P_{\uparrow}$ and $P_{\downarrow}$ states, respectively. (a--d) $\delta S_y^{\mathrm{even}}$, $\delta S_z^{\mathrm{even}}$, $\delta S_y^{\mathrm{odd}}$, and $\delta S_z^{\mathrm{odd}}$, respectively, for $\mathbf m\parallel +x$. Upon switching from $P_{\uparrow}$ to $P_{\downarrow}$, $\delta S_y^{\mathrm{even}}$ and $\delta S_z^{\mathrm{odd}}$ reverse sign, whereas $\delta S_z^{\mathrm{even}}$ and $\delta S_y^{\mathrm{odd}}$ remain unchanged. (e--g) $\delta S_y^{\mathrm{even}}$, $\delta S_z^{\mathrm{even}}$, and $\delta S_x^{\mathrm{odd}}$, respectively, for $\mathbf m\parallel +y$. Here, only $\delta S_y^{\mathrm{even}}$ reverses sign, while $\delta S_z^{\mathrm{even}}$ and $\delta S_x^{\mathrm{odd}}$ are unchanged. (h,i) $\delta S_y^{\mathrm{even}}$ and $\delta S_x^{\mathrm{odd}}$, respectively, for $\mathbf m\parallel +z$, both of which reverse sign between the two ferroelectric states.  }
\label{fig3}
\end{figure*}

\subsection{Ferroelectric-state and energy dependence of the Edelstein response}

We first examine the energy-dependent Edelstein response of the $P_{\uparrow}$ state. According to the symmetry analysis in Appendix~A, an electric field along $+x$ allows $\delta S_{y,z}^{\mathrm{even/odd}}$ for $\mathbf m\parallel+x$, $\delta S_{y,z}^{\mathrm{even}}$ and $\delta S_x^{\mathrm{odd}}$ for $\mathbf m\parallel+y$, and $\delta S_y^{\mathrm{even}}$ and $\delta S_x^{\mathrm{odd}}$ for $\mathbf m\parallel+z$. The corresponding energy dependences are shown in Fig.~\ref{fig2}. For $\mathbf m\parallel+x$ [Figs.~\ref{fig2}(a) and \ref{fig2}(b)], all four allowed components vary strongly with energy. For example, $\delta S_y^{\mathrm{even}}$ reaches approximately $0.12~\hbar$ near $E-E_F=-0.42$~eV and changes sign on the electron doping side, while $\delta S_z^{\mathrm{even}}$ decreases to about $-0.35~\hbar$ near $+0.49$~eV. The odd components are smaller, with $\delta S_y^{\mathrm{odd}}\simeq0.07~\hbar$ near $-0.32$~eV and $\delta S_z^{\mathrm{odd}}\simeq-0.10~\hbar$ near $-0.41$~eV. The element-resolved results in Fig.~\ref{fig11} show that the large negative $\delta S_z^{\mathrm{even}}$ near $+0.49$~eV is mainly associated with the Fe contribution, which reaches about $-0.25~\hbar$, while the Ga and Te contributions provide smaller corrections. This is consistent with the dominant Fe-$d$ character of the low-energy electronic structure shown in Fig.~\ref{fig1}.

A much stronger response appears for $\mathbf m\parallel+y$ [Figs.~\ref{fig2}(c) and \ref{fig2}(d)]. In particular, $\delta S_y^{\mathrm{even}}$ changes rapidly from a positive peak of nearly $2.0~\hbar$ around $E-E_F=+0.29$~eV to a negative value of about $-1.5~\hbar$ around $+0.34$~eV. Under the same reference field of $1$ V/\AA, the peak magnitude of nearly $2.0~\hbar$ exceeds the approximately $1.2~\hbar$ maximum reported for AB/BA-stacked bilayer 2$H$-VTe$_2$~\cite{Pan2026VTe2}. The element decomposition in Fig.~\ref{fig11}(e) further shows that this large variation is dominated by the Fe sites: the summed Fe contribution reaches about $1.1~\hbar$ near the positive peak, while the Te contribution is also sizable, approximately $0.6~\hbar$, and the Ga contribution is considerably smaller. The signs of the Fe and Te contributions subsequently reverse over a narrow energy interval, producing the pronounced sign change of the total response. The second even component, $\delta S_z^{\mathrm{even}}$, is much smaller but increases to approximately $0.3~\hbar$ near $+0.48$~eV. By comparison, the allowed odd component $\delta S_x^{\mathrm{odd}}$ remains below about $0.1~\hbar$ throughout the calculated energy window.

For $\mathbf m\parallel+z$ [Figs.~\ref{fig2}(e) and \ref{fig2}(f)], the reduced set of allowed components shows a more moderate energy dependence. The even response $\delta S_y^{\mathrm{even}}$ reaches about $0.13~\hbar$ near $-0.23$~eV and changes to approximately $-0.14~\hbar$ near $+0.47$~eV, whereas $\delta S_x^{\mathrm{odd}}$ is about $0.06~\hbar$ near $-0.38$~eV and develops a negative minimum of roughly $-0.1~\hbar$ around $+0.30$~eV. As shown in Figs.~\ref{fig11}(h) and \ref{fig11}(i), these responses again arise predominantly from the Fe contribution, with appreciable Te contributions and a smaller Ga contribution. The element-resolved behavior is therefore consistent with the Fe-$d$-dominated bands near the Fermi level, while the finite Ga and Te contributions indicate that the induced spin accumulation cannot be assigned to the Fe sublattice alone. Related sublattice-resolved calculations for monolayer Fe$_3$GeTe$_2$ and Fe$_3$GaTe$_2$ likewise show that current-induced spin--orbit torques contain strongly site-dependent and partially compensating contributions~\cite{Brizolla2026SOT}. The representative energies marked in Fig.~\ref{fig2} are also indicated on the corresponding band structures in Fig.~\ref{fig12}. Within the Kubo decomposition used here, the TR-even Edelstein response is predominantly an intraband Fermi-surface term, whereas the TR-odd response is interband and is enhanced near small spin--orbit-coupled band separations~\cite{Freimuth2014SOT,Johansson2024EdelsteinReview,Pan2026VTe2}. Further details of the layer-resolved results are provided in Appendix~L.

\begin{figure*}[htp]
\includegraphics[width=180mm]{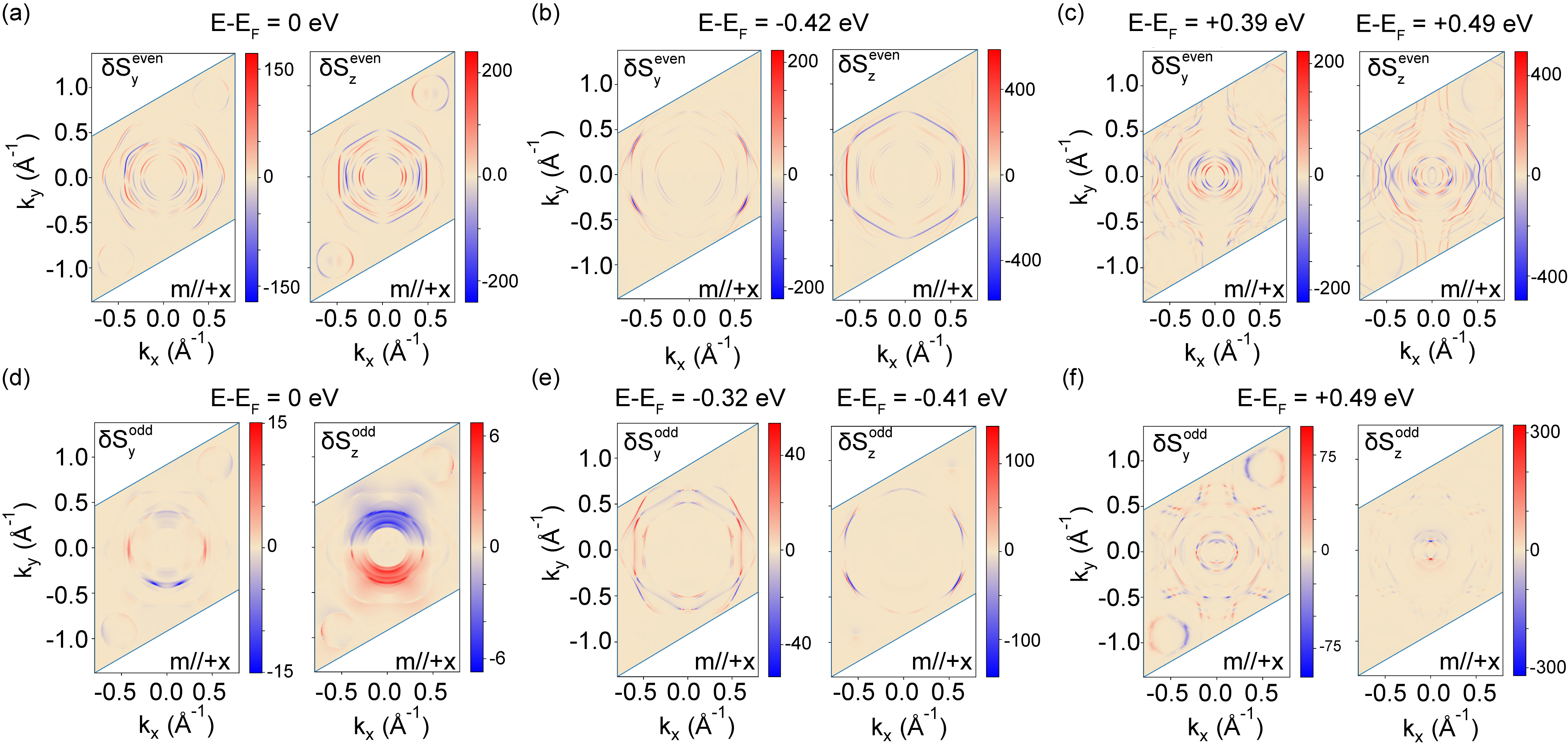}
\caption{ Momentum-resolved current-induced spin accumulation in $P_{\uparrow}$ state for $\mathbf m\parallel +x$ under an electric field applied along the $+x$ direction, with the reference field $E_x=1$~V/\AA. (a) $\delta S_y^{\mathrm{even}}(\mathbf k)$ and $\delta S_z^{\mathrm{even}}(\mathbf k)$ at the Fermi level, $E-E_F=0$. (b) Corresponding even components at $E-E_F=-0.42$~eV. (c) $\delta S_y^{\mathrm{even}}(\mathbf k)$ at $E-E_F=+0.39$~eV and $\delta S_z^{\mathrm{even}}(\mathbf k)$ at $E-E_F=+0.49$~eV. (d) $\delta S_y^{\mathrm{odd}}(\mathbf k)$ and $\delta S_z^{\mathrm{odd}}(\mathbf k)$ at $E-E_F=0$. (e) $\delta S_y^{\mathrm{odd}}(\mathbf k)$ at $E-E_F=-0.32$~eV and $\delta S_z^{\mathrm{odd}}(\mathbf k)$ at $E-E_F=-0.41$~eV. (f) Corresponding odd components at $E-E_F=+0.49$~eV. Red and blue denote positive and negative contributions, respectively. The representative energies are selected from the extrema of the energy-dependent responses shown in Fig.~\ref{fig2}. }
\label{fig4}
\end{figure*}

\begin{figure*}[]
\includegraphics[width=180mm]{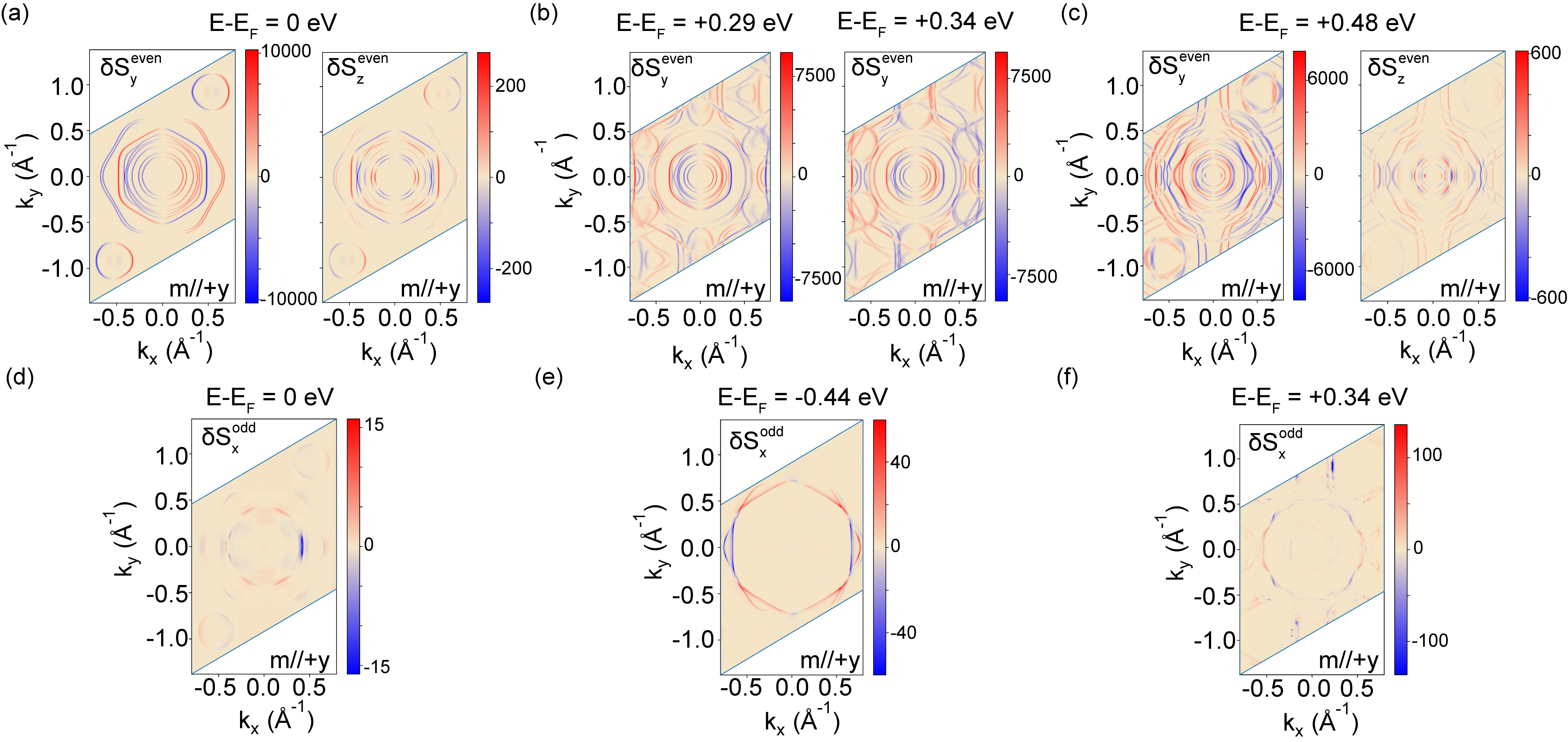}
\caption{ Momentum-resolved current-induced spin accumulation in $P_{\uparrow}$ state for $\mathbf m\parallel +y$ under an electric field applied along the $+x$ direction, with the reference field $E_x=1$~V/\AA. (a) $\delta S_y^{\mathrm{even}}(\mathbf k)$ and $\delta S_z^{\mathrm{even}}(\mathbf k)$ at the Fermi level, $E-E_F=0$. (b) $\delta S_y^{\mathrm{even}}(\mathbf k)$ at $E-E_F=+0.29$ and $+0.34$~eV. (c) $\delta S_y^{\mathrm{even}}(\mathbf k)$ and $\delta S_z^{\mathrm{even}}(\mathbf k)$ at $E-E_F=+0.48$~eV. (d--f) $\delta S_x^{\mathrm{odd}}(\mathbf k)$ at $E-E_F=0$, $-0.44$, and $+0.34$~eV, respectively. Red and blue denote positive and negative contributions, respectively. The representative energies are selected from the extrema of the energy-dependent responses shown in Fig.~\ref{fig2}. }
\label{fig5}
\end{figure*}

\begin{figure*}[t]
\includegraphics[width=103mm]{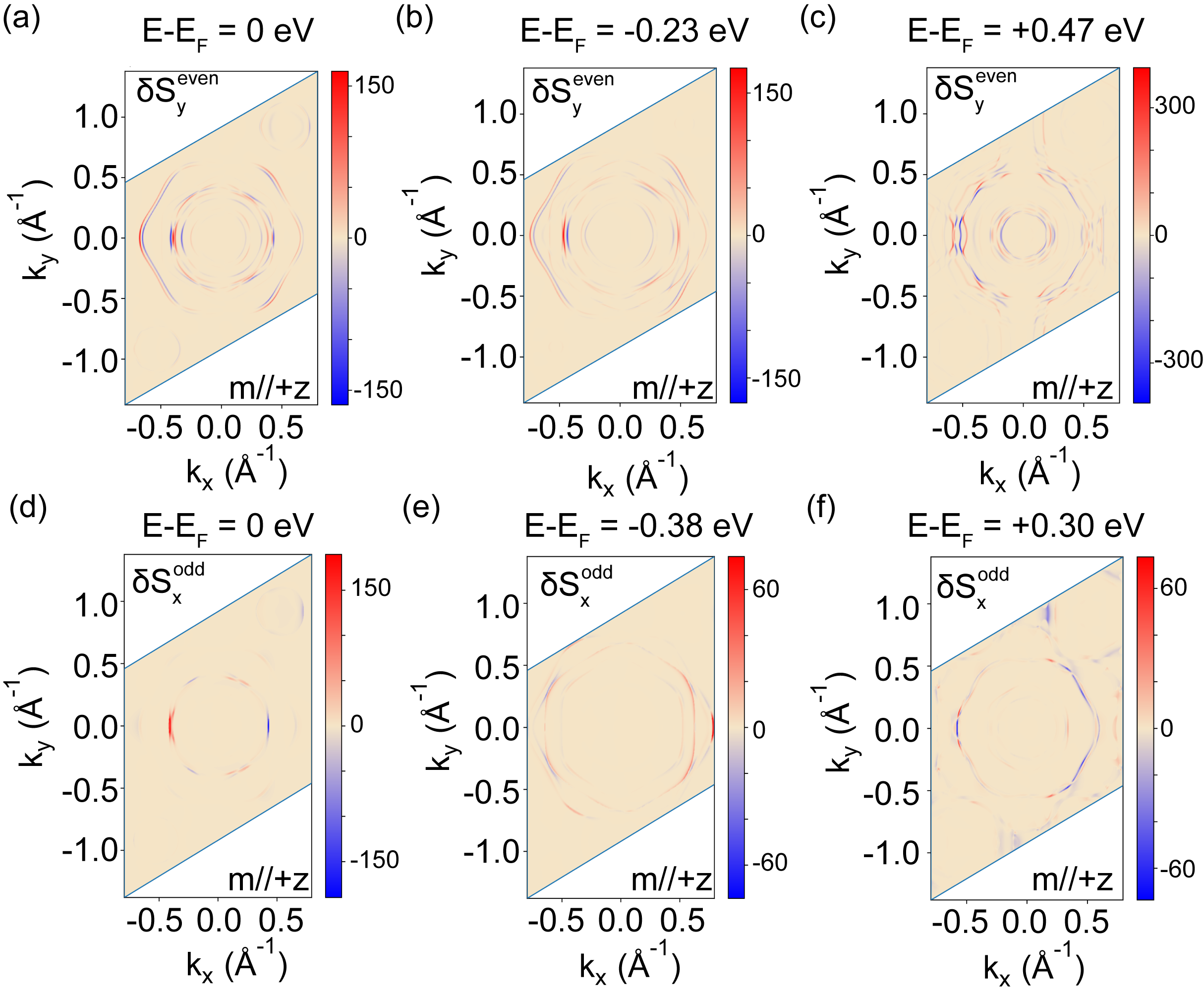}
\caption{ Momentum-resolved current-induced spin accumulation in $P_{\uparrow}$ state for $\mathbf m\parallel +z$ under an electric field applied along the $+x$ direction, with the reference field $E_x=1$~V/\AA. (a--c) $\delta S_y^{\mathrm{even}}(\mathbf k)$ at $E-E_F=0$, $-0.23$, and $+0.47$~eV, respectively. (d--f) $\delta S_x^{\mathrm{odd}}(\mathbf k)$ at $E-E_F=0$, $-0.38$, and $+0.30$~eV, respectively. Red and blue denote positive and negative contributions, respectively. The representative energies are selected from the extrema of the energy-dependent responses shown in Fig.~\ref{fig2}. }
\label{fig6}
\end{figure*}

We next examine how the Edelstein response changes upon reversing the ferroelectric polarization. As derived in Appendix~B, the two sliding ferroelectric states are structurally related by $M_z$. For a fixed in-plane magnetization, the corresponding magnetic states are connected by $M_z\mathcal T$, whereas for $\mathbf m\parallel z$ they are related directly by $M_z$. These symmetry relations impose component-dependent sign changes between $P_{\uparrow}$ and $P_{\downarrow}$ state. The calculated results in Fig.~\ref{fig3} follow these constraints over the entire energy range. For $\mathbf m\parallel+x$, $\delta S_y^{\mathrm{even}}$ and $\delta S_z^{\mathrm{odd}}$ reverse sign between the two ferroelectric states, while $\delta S_z^{\mathrm{even}}$ and $\delta S_y^{\mathrm{odd}}$ remain unchanged [Figs.~\ref{fig3}(a)-\ref{fig3}(d)]. For $\mathbf m\parallel+y$, only $\delta S_y^{\mathrm{even}}$ changes sign, whereas $\delta S_z^{\mathrm{even}}$ and $\delta S_x^{\mathrm{odd}}$ are preserved [Figs.~\ref{fig3}(e)-\ref{fig3}(g)]. When $\mathbf m\parallel+z$, both allowed components, $\delta S_y^{\mathrm{even}}$ and $\delta S_x^{\mathrm{odd}}$, reverse sign upon switching the ferroelectric state [Figs.~\ref{fig3}(h)-\ref{fig3}(i)]. Thus, ferroelectric switching does not simply reverse the entire Edelstein response, but selectively controls its spin-polarization components according to the magnetization direction. For inversion-related polarization domains of a nonmagnetic ferroelectric metal, the leading Edelstein coefficient or spin texture typically reverses with the polarization~\cite{Pan2026PtBi2,Rinaldi2018GeTe}. Here, the additional magnetic order makes the domain transformation magnetization dependent and therefore produces component-selective, rather than uniform, sign reversals. The coexistence of ferroelectric and magnetic degrees of freedom thus provides an additional symmetry control over the current-induced spin polarization. Since the responses of the two ferroelectric states are fully related by symmetry, we focus below on $P_{\uparrow}$ state to analyze the microscopic momentum-space origin of the Edelstein response.

\subsection{Momentum-space origin of the Edelstein response}

To resolve the microscopic origin of the energy-dependent Edelstein response, we calculate the momentum-resolved spin accumulation for $P_{\uparrow}$ state at the representative energies selected in Fig.~\ref{fig2}. Since the TR-even contribution is mainly intraband, its momentum-resolved distribution follows the constant-energy contours at the selected chemical potential. The TR-odd contribution is instead associated with interband processes and is therefore more sensitive to small energy separations between spin--orbit-coupled bands~\cite{Freimuth2014SOT,Johansson2024EdelsteinReview,Pan2026VTe2}. The momentum-space patterns are further restricted by the magnetic symmetries derived in Appendix G. For $\mathbf m\parallel+x$, the allowed components are symmetric with respect to the $k_x=0$ line, whereas for $\mathbf m\parallel+y$ and $\mathbf m\parallel+z$ the corresponding mirror axis is $k_y=0$. These relations are directly visible in Figs.~\ref{fig4}--\ref{fig6}.

For $\mathbf m\parallel+x$, the two TR-even components at the Fermi level in Fig.~\ref{fig4}(a) are distributed mainly along several $\Gamma$-centered contours. Both $\delta S_y^{\mathrm{even}}(\mathbf k)$ and $\delta S_z^{\mathrm{even}}(\mathbf k)$ contain positive and negative arc-like segments, leading to partial cancellation after integration over the Brillouin zone. At $E-E_F=-0.42$~eV [Fig.~\ref{fig4}(b)], the response is concentrated on fewer, well-defined large contours; in particular, $\delta S_z^{\mathrm{even}}(\mathbf k)$ develops a pronounced hexagonal arc-like distribution. Moving to the electron-doping side, $\delta S_y^{\mathrm{even}}(\mathbf k)$ at $+0.39$~eV and $\delta S_z^{\mathrm{even}}(\mathbf k)$ at $+0.49$~eV [Fig.~\ref{fig4}(c)] show considerably more fragmented and sign-alternating patterns. This evolution is consistent with the dense Fe-$d$ multiband structure intersected by these energies in Fig.~\ref{fig12}(a). The TR-odd distributions show a different character. At the Fermi level [Fig.~\ref{fig4}(d)], $\delta S_y^{\mathrm{odd}}(\mathbf k)$ is mainly confined to arc-like segments around the central contour, whereas $\delta S_z^{\mathrm{odd}}(\mathbf k)$ forms broader positive and negative regions around $\Gamma$. At $-0.32$ and $-0.41$~eV [Fig.~\ref{fig4}(e)], the odd response becomes concentrated on restricted portions of the large constant-energy contours. At $+0.49$~eV [Fig.~\ref{fig4}(f)], both odd components develop additional localized arc-like structures, including contributions near the central and outer regions of the Brillouin zone. 

The redistribution is particularly pronounced for $\mathbf m\parallel+y$ in Fig.~\ref{fig5}. At the Fermi level [Fig.~\ref{fig5}(a)], $\delta S_y^{\mathrm{even}}(\mathbf k)$ exhibits several concentric $\Gamma$-centered arc-like contours with alternating signs, while $\delta S_z^{\mathrm{even}}(\mathbf k)$ shows a similar but much weaker distribution. The most striking change occurs in Fig.~\ref{fig5}(b). At $E-E_F=+0.29$ and $+0.34$~eV, which correspond to the large positive and negative extrema of $\delta S_y^{\mathrm{even}}$ in Fig.~\ref{fig2}(c), the relative weights of the positive and negative arc-like contributions are substantially reorganized over an energy interval of only $0.05$~eV. The resulting change in their balance accounts for the rapid sign reversal of the integrated response. At $+0.48$~eV [Fig.~\ref{fig5}(c)], $\delta S_y^{\mathrm{even}}(\mathbf k)$ develops an extended sign-alternating network of arc-like features, whereas $\delta S_z^{\mathrm{even}}(\mathbf k)$ remains more localized near the central contours. These complex distributions are consistent with the closely spaced Fe-$d$ bands in Fig.~\ref{fig12}(b). The TR-odd $\delta S_x^{\mathrm{odd}}(\mathbf k)$ is much weaker at the Fermi level [Fig.~\ref{fig5}(d)]. At $-0.44$~eV [Fig.~\ref{fig5}(e)], the response becomes concentrated on an almost continuous large contour with alternating positive and negative arc-like segments, whereas at $+0.34$~eV [Fig.~\ref{fig5}(f)] the distribution breaks into more localized features. All of these patterns are symmetric about $k_y=0$, as required by $M_x\mathcal T$.

For $\mathbf m\parallel+z$, only $\delta S_y^{\mathrm{even}}$ and $\delta S_x^{\mathrm{odd}}$ remain allowed, and their momentum-space distributions are correspondingly simpler (Fig.~\ref{fig6}). The even response is mainly carried by several $\Gamma$-centered contours at $-0.23$~eV [Fig.~\ref{fig6}(b)] and becomes more structured at $+0.47$~eV [Fig.~\ref{fig6}(c)], while the odd component develops localized sign-changing features at $-0.38$ and $+0.30$~eV [Fig.~\ref{fig6}(e) and ~\ref{fig6}(f)]. The results in Figs.~\ref{fig4}--\ref{fig6} show that changing the magnetization direction modifies not only the symmetry-allowed spin components but also the balance between positive and negative intraband contributions and the strength of interband processes.

\subsection{Magnetization-angle dependence of the Edelstein response}

\begin{figure}[t]
\includegraphics[width=85mm]{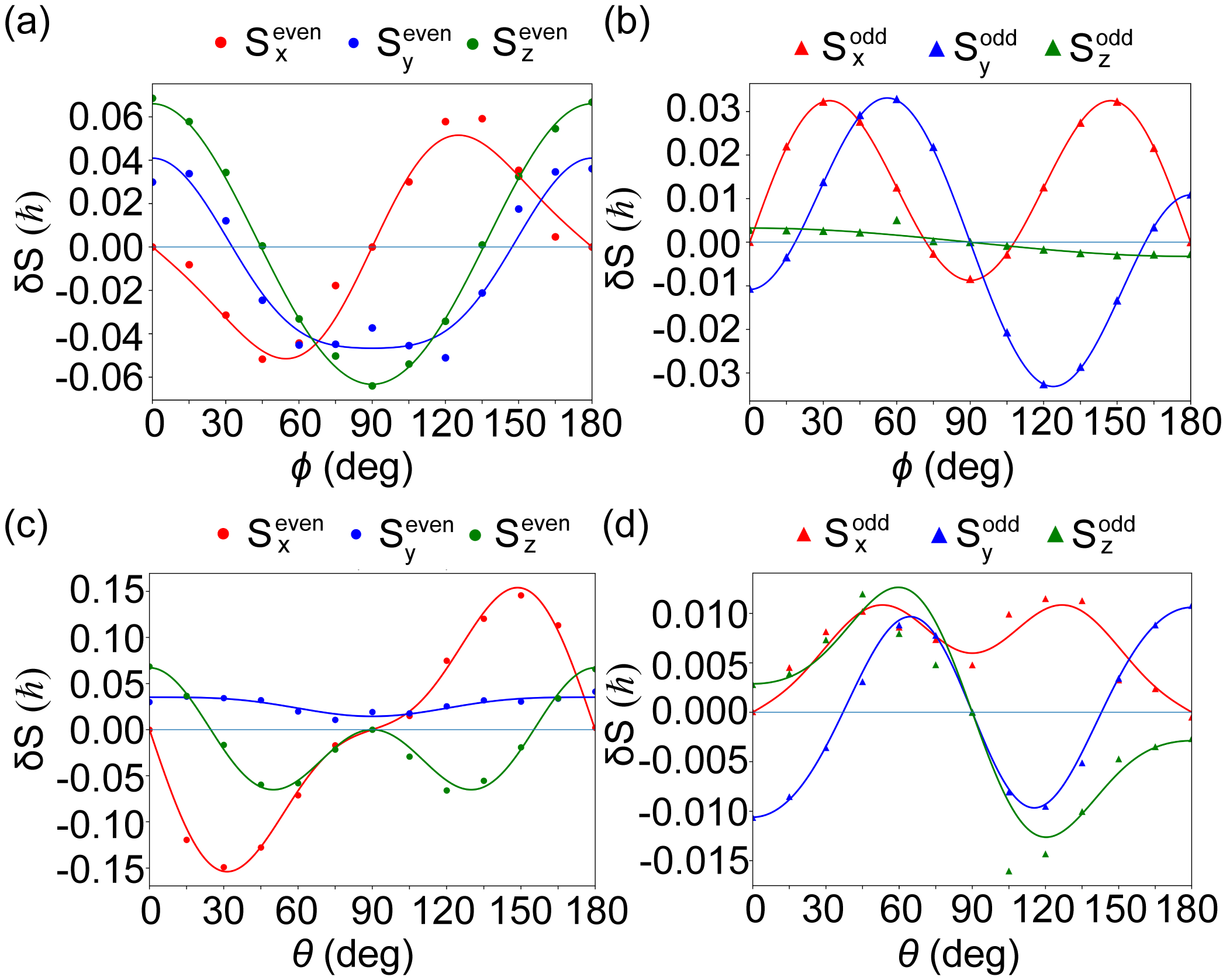}
\caption{ Angular dependence of the current-induced spin accumulation in $P_{\uparrow}$ state at the Fermi level under an electric field $E_x=1$~V/\AA. (a,b) Time-reversal-even and -odd components, respectively, for magnetization rotating in the $xy$ plane, $\mathbf m=(\cos\phi,\sin\phi,0)$. The solid lines are fits to $\delta S_x^{\mathrm{even}}=A_2\sin2\phi+A_4\sin4\phi$, $\delta S_y^{\mathrm{even}}=B_0+B_2\cos2\phi+B_4\cos4\phi$, and $\delta S_z^{\mathrm{even}}=C_2\cos2\phi+C_4\cos4\phi$ for the even response, and $\delta S_x^{\mathrm{odd}}=D_1\sin\phi+D_3\sin3\phi+D_5\sin5\phi$, $\delta S_y^{\mathrm{odd}}=D_1\cos\phi+E_3\cos3\phi-D_5\cos5\phi$, and $\delta S_z^{\mathrm{odd}}=F_1\cos\phi+F_5\cos5\phi$ for the odd response. (c,d) Corresponding even and odd components for magnetization rotating in the $xz$ plane, $\mathbf m=(\cos\theta,0,\sin\theta)$. The fitting forms are $\delta S_x^{\mathrm{even}}=G_2\sin2\theta+G_4\sin4\theta$, $\delta S_y^{\mathrm{even}}=H_0+H_2\cos2\theta+H_4\cos4\theta$, $\delta S_z^{\mathrm{even}}=I_2\cos^2\theta+I_4\cos^4\theta$, $\delta S_x^{\mathrm{odd}}=J_1\sin\theta+J_3\sin3\theta+J_5\sin5\theta$, $\delta S_y^{\mathrm{odd}}=K_1\cos\theta+K_3\cos3\theta+K_5\cos5\theta$, and $\delta S_z^{\mathrm{odd}}=L_1\cos\theta+L_3\cos3\theta+L_5\cos5\theta$. Red, blue, and green denote the $x$, $y$, and $z$ spin components, respectively. Circles and triangles represent the calculated even and odd spin accumulations, respectively. The corresponding fitting coefficients are summarized in Table~\ref{table2}. }
\label{fig7}
\end{figure}

The three principal magnetization directions considered above represent only the high-symmetry limits of the magnetic anisotropy. Figure~\ref{fig7} shows the continuous evolution of the Edelstein response as the magnetization is rotated in the $xy$ and $xz$ planes. The symmetry-adapted angular forms are derived in Appendix~C, and the corresponding fitting coefficients are listed in Table~\ref{table2}. For the $xy$-plane rotation, $\mathbf m=(\cos\phi,\sin\phi,0)$, the calculated responses in Figs.~\ref{fig7}(a) and \ref{fig7}(b) evolve smoothly between the $\mathbf m\parallel+x$ and $\mathbf m\parallel+y$ limits at $\phi=0^\circ$ and $90^\circ$, respectively. In the TR-even sector, \(\delta S_x^{\mathrm{even}}\) vanishes along both high-symmetry directions but becomes finite at intermediate angles, reaching a maximum magnitude of about \(0.06~\hbar\). Meanwhile, \(\delta S_y^{\mathrm{even}}\) evolves from approximately \(+0.04~\hbar\) to \(-0.05~\hbar\), while \(\delta S_z^{\mathrm{even}}\) changes from about \(+0.066~\hbar\) at \(\phi=0^\circ\) to \(-0.063~\hbar\) at \(\phi=90^\circ\). The odd response shows a more pronounced departure from a single sinusoidal form: $\delta S_x^{\mathrm{odd}}$ and $\delta S_y^{\mathrm{odd}}$ reach approximately $0.03~\hbar$, whereas $\delta S_z^{\mathrm{odd}}$ remains much smaller, of order $3\times10^{-3}~\hbar$. At $\phi=90^\circ$, $\delta S_y^{\mathrm{odd}}$ and $\delta S_z^{\mathrm{odd}}$ vanish, leaving $\delta S_x^{\mathrm{odd}}$ as the only allowed odd component, consistent with the $\mathbf m\parallel+y$ selection rule derived in Appendix~A.

\begin{figure*}[t]
\includegraphics[width=175mm]{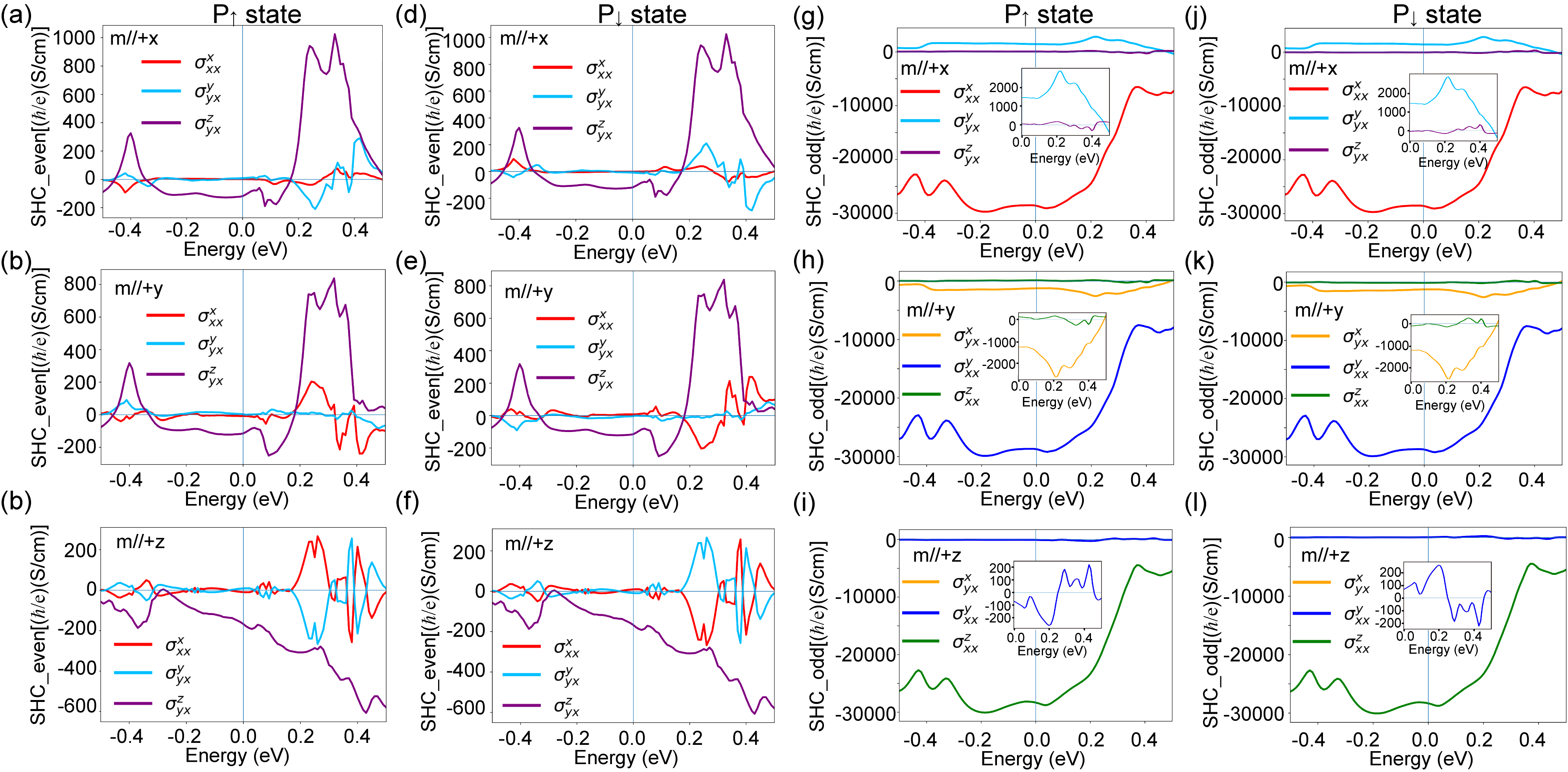}
\caption{ Calculated energy-dependent spin-conductivity tensor components for $P_{\uparrow}$ and $P_{\downarrow}$ states and different magnetization directions. Panels (a--c) and (d--f) show the time-reversal-even spin-conductivity components for $P_{\uparrow}$ and $P_{\downarrow}$ states, respectively, whereas panels (g--i) and (j--l) show the corresponding time-reversal-odd contributions. Within each group, the magnetization is oriented along $\mathbf m\parallel +x$, $+y$, and $+z$ from top to bottom. The electric field is applied along the $+x$ direction. For the even response, the displayed components are $\sigma_{xx}^{x}$, $\sigma_{yx}^{y}$, and $\sigma_{yx}^{z}$ for all three magnetization directions. For the odd response with $\mathbf m\parallel +x$, the displayed components are $\sigma_{xx}^{x}$, $\sigma_{yx}^{y}$, and $\sigma_{yx}^{z}$, whereas for $\mathbf m\parallel +y$ and $\mathbf m\parallel +z$, they are $\sigma_{yx}^{x}$, $\sigma_{xx}^{y}$, and $\sigma_{xx}^{z}$. The insets in panels (g--l) enlarge the comparatively smaller odd components: $\sigma_{yx}^{y}$ and $\sigma_{yx}^{z}$ for $\mathbf m\parallel +x$, $\sigma_{yx}^{x}$ and $\sigma_{xx}^{z}$ for $\mathbf m\parallel +y$, and $\sigma_{yx}^{x}$ and $\sigma_{xx}^{y}$ for $\mathbf m\parallel +z$. The vertical blue lines indicate the Fermi level. The conductivities are given in units of $(\hbar/e)(\mathrm{S/cm})$ after applying the effective-thickness correction for the two-dimensional slab. }
\label{fig8}
\end{figure*}

A stronger angular anisotropy appears for rotation in the $xz$ plane, $\mathbf m=(\cos\theta,0,\sin\theta)$ [Figs.~\ref{fig7}(c) and \ref{fig7}(d)]. In particular, \(\delta S_x^{\mathrm{even}}\) exhibits the strongest angular variation, reaching a magnitude of approximately \(0.15~\hbar\) near \(150^\circ\), while vanishing for both \(\mathbf m\parallel x\) and \(\mathbf m\parallel z\). The $\delta S_z^{\mathrm{even}}$ component also varies nonmonotonically: it starts from about $0.067~\hbar$ at $\theta=0^\circ$, changes sign at intermediate angles, and vanishes when $\theta=90^\circ$. By comparison, $\delta S_y^{\mathrm{even}}$ remains positive and varies more weakly, between approximately $0.014~\hbar$ and $0.035~\hbar$. The three TR-odd components are smaller, with characteristic magnitudes of order $10^{-2}~\hbar$, but their angular profiles also contain appreciable higher-order contributions. At $\theta=90^\circ$, corresponding to $\mathbf m\parallel+z$, all components vanish except $\delta S_y^{\mathrm{even}}$ and $\delta S_x^{\mathrm{odd}}$, recovering the $3m'$ symmetry constraint. In both rotation planes, the symmetry-constrained fits require higher-order harmonics beyond the leading terms, consistent with the angular dependence reported for $C_{3v}$-symmetric bilayer 2$H$-\ce{VTe2} and with recent first-principles calculations of current-induced torques in the same material family~\cite{Pan2026VTe2,Brizolla2026SOT}.

\begin{table}[htbp]
\centering
\caption{
Fitting coefficients for the angular dependence of the Edelstein spin
accumulation for magnetization rotations in the $xy$ and $xz$ planes.
The coefficients are given in units of $~\hbar$/cell for
$E_x=1$~V/\AA.
}
\begin{tabular}{cccc}
\hline\hline
\multicolumn{2}{c}{$xy$-plane rotation} &
\multicolumn{2}{c}{$xz$-plane rotation} \\
Coefficient & Value & Coefficient & Value \\
\hline
$A_2$ & $-4.7906\times10^{-2}$ &
$G_2$ & $-1.2700\times10^{-1}$ \\

$A_4$ & $+1.0135\times10^{-2}$ &
$G_4$ & $-5.0466\times10^{-2}$ \\

$B_0$ & $-1.2195\times10^{-2}$ &
$H_0$ & $+2.7242\times10^{-2}$ \\

$B_2$ & $+4.3815\times10^{-2}$ &
$H_2$ & $+1.0387\times10^{-2}$ \\

$B_4$ & $+9.3155\times10^{-3}$ &
$H_4$ & $-2.4334\times10^{-3}$ \\

$C_2$ & $+6.4672\times10^{-2}$ &
$I_2$ & $-3.1585\times10^{-1}$ \\

$C_4$ & $+1.3416\times10^{-3}$ &
$I_4$ & $+3.8295\times10^{-1}$ \\

$D_1$ & $+1.5066\times10^{-2}$ &
$J_1$ & $+1.0215\times10^{-2}$ \\

$D_3$ & $+2.4403\times10^{-2}$ &
$J_3$ & $+2.4072\times10^{-3}$ \\

$D_5$ & $+5.8045\times10^{-4}$ &
$J_5$ & $-1.8322\times10^{-3}$ \\

$E_3$ & $-2.5319\times10^{-2}$ &
$K_1$ & $-2.4346\times10^{-3}$ \\

$F_1$ & $+3.2510\times10^{-3}$ &
$K_3$ & $-9.7372\times10^{-3}$ \\

$F_5$ & $-6.7112\times10^{-6}$ &
$K_5$ & $+1.5679\times10^{-3}$ \\

     &                              &
$L_1$ & $+8.7062\times10^{-3}$ \\

     &                              &
$L_3$ & $-7.4603\times10^{-3}$ \\

     &                              &
$L_5$ & $+1.6314\times10^{-3}$ \\
\hline\hline
\end{tabular}
\label{table2}
\end{table}

\subsection{Ferroelectric-state and energy dependence of the spin-conductivity response}

The Edelstein effect characterizes the nonequilibrium spin accumulation generated by an electric field, while spin-current generation provides a complementary channel for charge-to-spin conversion. In particular, a transverse spin current transports spin angular momentum away from the charge-current direction and can generate spin--orbit torque when absorbed by a magnetic order~\cite{Sinova2015SHE,Manchon2019SOT}. We therefore turn to the energy-dependent spin-conductivity tensor of bilayer \ce{Fe3GaTe2}.

For the two ferroelectric states, the spin-conductivity tensors are constrained by the symmetry relations derived in Appendix~B. For fixed in-plane magnetization, $\mathbf m\parallel x$ or $y$,  $P_{\uparrow}$ and $P_{\downarrow}$ states are connected by $M_z\mathcal T$. Accordingly, the TR-even components with spin polarization along $x$ or $y$ change sign upon ferroelectric switching, while the $z$-polarized components are preserved. In the TR-odd sector, the $x$- and $y$-polarized components remain unchanged, whereas the $z$-polarized components reverse sign. For $\mathbf m\parallel z$, the two states are related by $M_z$, under which the $x$- and $y$-polarized components change sign and the $z$-polarized components remain invariant for both TR-even and TR-odd responses.

Figure~\ref{fig8} shows that these symmetry relations are satisfied throughout the calculated energy window. In the TR-even sector [Figs.~\ref{fig8}(a)--\ref{fig8}(f)], $\sigma_{xx}^{x}$ and $\sigma_{yx}^{y}$ reverse sign between $P_{\uparrow}$ and $P_{\downarrow}$ states, whereas $\sigma_{yx}^{z}$ is unchanged for all three principal magnetization directions. The response is strongly energy dependent. For $\mathbf m\parallel+x$, the transverse $\sigma_{yx}^{z}$ component reaches nearly $10^{3}(~\hbar/e)(\mathrm{S/cm})$ on the electron-doping side, while $\sigma_{xx}^{x}$ and $\sigma_{yx}^{y}$ remain considerably smaller. A similar large positive $\sigma_{yx}^{z}$, approaching $8\times10^{2}(\hbar/e)(\mathrm{S/cm})$, appears for $\mathbf m\parallel+y$. For $\mathbf m\parallel+z$, the same component changes sign and reaches approximately $-6\times10^{2}(\hbar/e)(\mathrm{S/cm})$ at positive energies. The pronounced variation of both magnitude and sign with magnetization direction reflects the strong anisotropy of the spin-current response.

The TR-odd sector [Figs.~\ref{fig8}(g)--\ref{fig8}(l)] has a markedly different scale. For in-plane magnetization, $\sigma_{xx}^{x,\mathrm{odd}}$ for $\mathbf m\parallel+x$ and $\sigma_{xx}^{y,\mathrm{odd}}$ for $\mathbf m\parallel+y$ are unchanged between the two ferroelectric states, whereas the corresponding $z$-polarized components reverse sign. For $\mathbf m\parallel+z$, $\sigma_{xx}^{z,\mathrm{odd}}$ remains unchanged, while the $x$- and $y$-polarized components reverse sign, in agreement with the direct $M_z$ relation. A particularly notable feature is the large longitudinal response: $\sigma_{xx}^{x,\mathrm{odd}}$, $\sigma_{xx}^{y,\mathrm{odd}}$, and $\sigma_{xx}^{z,\mathrm{odd}}$ all reach magnitudes of order $3\times10^{4}(\hbar/e)(\mathrm{S/cm})$ over a broad energy range, substantially exceeding the transverse odd components. Moreover, the three longitudinal curves show very similar energy dependences and comparable magnitudes despite the different magnetization directions.

This near-equality is consistent with the intraband Fermi-surface form of the TR-odd spin conductivity in the constant-broadening Kubo formalism~\cite{Freimuth2014SOT,Gradhand2012BerryCurvature}. (The distinct broadening dependences of the interband and intraband sectors are verified in Appendix K and Fig.~\ref{fig13}.) In the
small-broadening limit, the dominant longitudinal contribution can be
written schematically as
\begin{equation}
\sigma_{xx}^{\alpha,\mathrm{odd}}
\propto
\frac{1}{\Gamma}
\sum_{n\mathbf{k}}
\delta(\varepsilon_{n\mathbf{k}}-E_F)
v_{x,n\mathbf{k}}^{2}
\langle S_{\alpha}\rangle_{n\mathbf{k}},
\label{eq:odd_longitudinal}
\end{equation}
where $v_{x,n\mathbf{k}}$ is the band velocity and
$\langle S_{\alpha}\rangle_{n\mathbf{k}}$ is the spin expectation
value of the corresponding Bloch state. The response is therefore
dominated by spin-polarized states at the Fermi surface. Since
\ce{Fe3GaTe2} is a ferromagnetic metal with sizable exchange
splitting, the spin expectation value of the relevant states tends,
to lowest order, to follow the magnetization direction,
\begin{equation}
\langle\mathbf S\rangle_{n\mathbf{k}}
\simeq
s_{n\mathbf{k}}\mathbf m .
\end{equation}
Substituting this form into Eq.~\eqref{eq:odd_longitudinal} gives
\begin{equation}
\begin{aligned}
\sigma_{xx}^{\alpha,\mathrm{odd}}
&\simeq
C_x(E_F)m_{\alpha},\\
C_x(E_F)
&\propto
\frac{1}{\Gamma}
\sum_{n\mathbf{k}}
\delta(\varepsilon_{n\mathbf{k}}-E_F)
s_{n\mathbf{k}}v_{x,n\mathbf{k}}^2 .
\end{aligned}
\label{eq:odd_follow_m}
\end{equation}
Accordingly,
\begin{equation}
\sigma_{xx}^{x,\mathrm{odd}}
(\mathbf m\parallel x)
\approx
\sigma_{xx}^{y,\mathrm{odd}}
(\mathbf m\parallel y)
\approx
\sigma_{xx}^{z,\mathrm{odd}}
(\mathbf m\parallel z),
\label{eq:odd_approx_relation}
\end{equation}
which accounts for the closely resembling longitudinal curves in
Figs.~\ref{fig8}(g)--\ref{fig8}(i) and
\ref{fig8}(j)--\ref{fig8}(l). Small differences among them arise
from spin--orbit-induced anisotropy and higher-order dependence on
the magnetization direction.

\begin{figure}[hbtp]
\includegraphics[width=85mm]{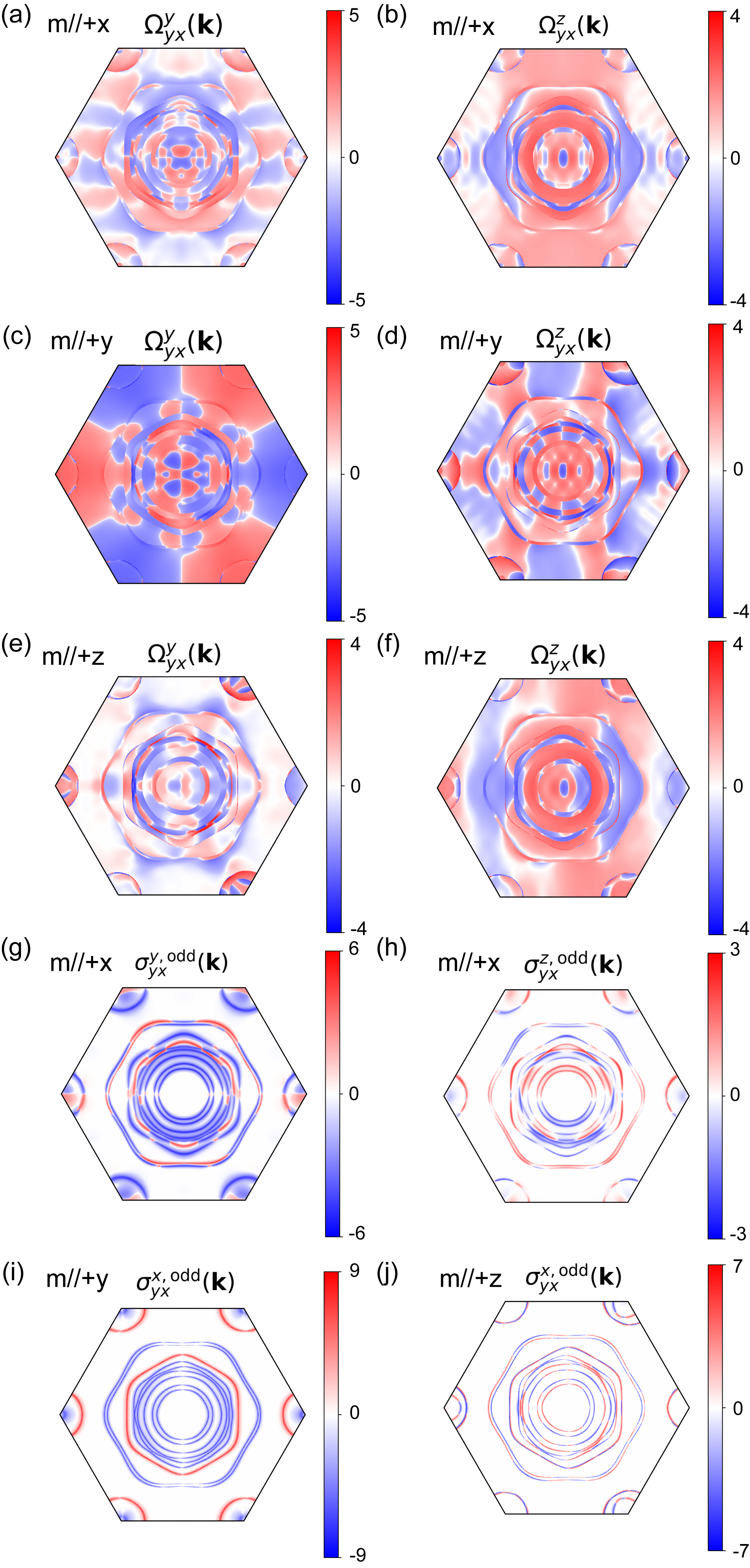}
\caption{ Momentum-resolved spin Hall response at the Fermi level. Panels (a)--(f) show the spin Berry curvature associated with the TR-even interband intrinsic response. (a,b) $\Omega_{yx}^{y}(\mathbf k)$ and $\Omega_{yx}^{z}(\mathbf k)$ for $\mathbf m\parallel+x$; (c,d) corresponding results for $\mathbf m\parallel+y$; and (e,f) corresponding results for $\mathbf m\parallel+z$. Panels (g)--(j) show the momentum-resolved TR-odd intraband spin conductivity. (g,h) $\sigma_{yx}^{y,\mathrm{odd}}(\mathbf k)$ and $\sigma_{yx}^{z,\mathrm{odd}}(\mathbf k)$ for $\mathbf m\parallel+x$; (i) $\sigma_{yx}^{x,\mathrm{odd}}(\mathbf k)$ for $\mathbf m\parallel+y$; and (j) $\sigma_{yx}^{x,\mathrm{odd}}(\mathbf k)$ for $\mathbf m\parallel+z$. Red and blue indicate positive and negative contributions, respectively, while the hexagonal boundary marks the first Brillouin zone. The spin Berry curvature is given in units of \(\AA^2\). }
\label{fig9}
\end{figure}

\subsection{Momentum-space origin of the spin Hall response}

To clarify the microscopic origin of the transverse spin-current response in Fig.~\ref{fig8}, we further examine its momentum-space distribution at the Fermi level. The two time-reversal sectors are represented by different momentum-resolved quantities because they originate from different parts of the Kubo response. The TR-even intrinsic contribution is interband in nature and can be expressed in terms of the spin Berry curvature $\Omega_{yx}^{\alpha}(\mathbf{k})$ \cite{Sinova2015SHE,Gradhand2012BerryCurvature}, whereas the TR-odd contribution is dominated by intraband Fermi-surface transport and is therefore shown directly as the momentum-resolved spin conductivity $\sigma_{yx}^{\alpha,\mathrm{odd}}(\mathbf{k})$. The former identifies the regions in momentum space where virtual interband transitions generate the intrinsic spin Hall response, while the latter identifies the Fermi-surface regions responsible for the odd transverse spin current.

Their momentum-space distributions are further constrained by the magnetic symmetries derived in Appendix G; Appendix H summarizes the Brillouin-zone mapping and plotting representation used in Figs.~\ref{fig9} and \ref{fig15}. For $\mathbf m\parallel+x$, the unitary mirror $M_x$ maps $(k_x,k_y)$ to $(-k_x,k_y)$, so that both the TR-even $\Omega_{yx}^{y,z}(\mathbf k)$ and the allowed TR-odd transverse components are symmetric with respect to the $k_x=0$ line. For $\mathbf m\parallel+y$ and $\mathbf m\parallel+z$, $M_x\mathcal T$ instead maps $(k_x,k_y)$ to $(k_x,-k_y)$, giving a mirror axis at $k_y=0$ for both sectors. The calculated patterns in Fig.~\ref{fig9} follow these relations. For $\mathbf m\parallel+z$, the additional $C_{3z}$ symmetry acts on the complete rank-three spin-conductivity tensor and therefore does not require each individual component to display an independent threefold pattern. These momentum-space constraints are the same for the TR-even intrinsic and TR-odd intraband responses, despite their different microscopic origins.

\begin{figure*}[t]
\includegraphics[width=170mm]{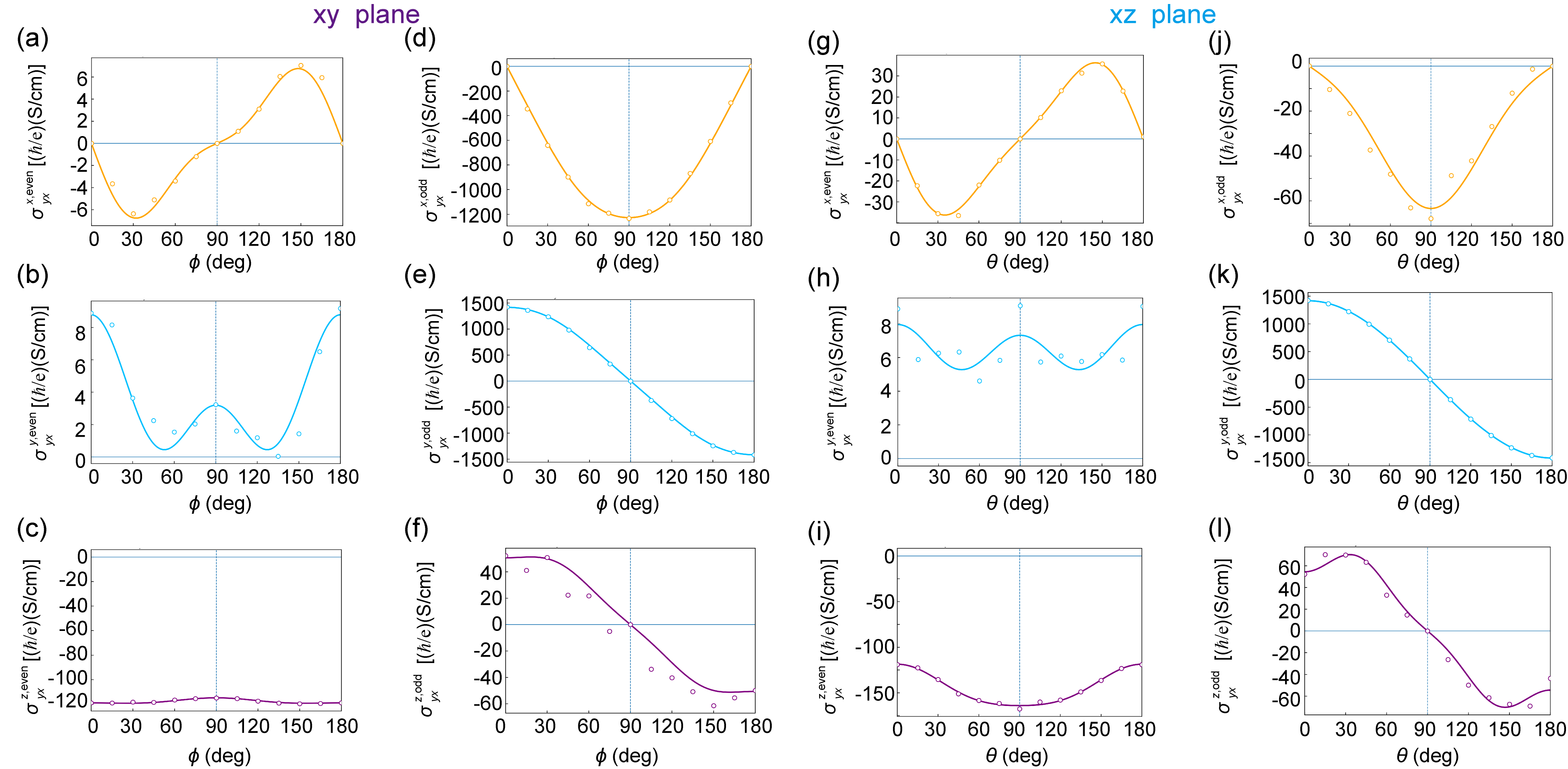}
\caption{ Magnetization-angle dependence of the transverse spin Hall conductivity in $P_{\uparrow}$ state at the Fermi level. Panels (a)--(c) and (d)--(f) show the time-reversal-even and time-reversal-odd components, respectively, for magnetization rotating in the $xy$ plane, $\mathbf m=(\cos\phi,\sin\phi,0)$. Panels (g)--(i) and (j)--(l) show the corresponding components for rotation in the $xz$ plane, $\mathbf m=(\cos\theta,0,\sin\theta)$. Within each group, the $x$-, $y$-, and $z$-spin-polarized transverse components $\sigma_{yx}^{x}$, $\sigma_{yx}^{y}$, and $\sigma_{yx}^{z}$ are shown from top to bottom. Open circles denote the calculated values, and solid curves are symmetry-constrained fits. For either rotation plane, with $\lambda=\phi$ for the $xy$ plane and $\lambda=\theta$ for the $xz$ plane, the TR-even components are fitted to $\sigma_{yx}^{x,\mathrm{even}} =A_2\sin2\lambda+A_4\sin4\lambda$, $\sigma_{yx}^{y,\mathrm{even}} =B_0+B_2\cos2\lambda+B_4\cos4\lambda$, and $\sigma_{yx}^{z,\mathrm{even}} =C_0+C_2\cos2\lambda+C_4\cos4\lambda$, whereas the TR-odd components are fitted to $\sigma_{yx}^{x,\mathrm{odd}} =D_1\sin\lambda+D_3\sin3\lambda+D_5\sin5\lambda$, $\sigma_{yx}^{y,\mathrm{odd}} =E_1\cos\lambda+E_3\cos3\lambda+E_5\cos5\lambda$, and $\sigma_{yx}^{z,\mathrm{odd}} =F_1\cos\lambda+F_3\cos3\lambda+F_5\cos5\lambda$. The fitting coefficients for the two rotation planes are summarized in Table~\ref{tab:SHC_angle_fit}. The vertical dashed lines mark $90^\circ$, corresponding to $\mathbf m\parallel+y$ for the $xy$ rotation and $\mathbf m\parallel+z$ for the $xz$ rotation. The conductivities are given in units of $(\hbar/e)(\mathrm{S/cm})$ after applying the effective-thickness correction for the two-dimensional slab. }
\label{fig10}
\end{figure*}

The TR-even spin Berry curvature in Figs.~\ref{fig9}(a)--\ref{fig9}(f) changes markedly with both spin polarization and magnetization direction. For $\mathbf m\parallel+x$, $\Omega_{yx}^{y}(\mathbf k)$ [Fig.~\ref{fig9}(a)] exhibits sign-alternating arcs and a multi-lobed structure around $\Gamma$, surrounded by a broad negative annular region and additional boundary contributions. In contrast, $\Omega_{yx}^{z}(\mathbf k)$ [Fig.~\ref{fig9}(b)] shows a more regular radial pattern, with a positive central region followed by alternating narrow rings and negative weight toward the zone edge. For $\mathbf m\parallel+y$, the SBC is substantially redistributed. $\Omega_{yx}^{y}(\mathbf k)$ [Fig.~\ref{fig9}(c)] contains comparable positive and negative contributions over both the central and outer regions of the Brillouin zone. These contributions largely cancel upon momentum integration, consistent with the relatively small $\sigma_{yx}^{y,\mathrm{even}}$ at the Fermi level in Fig.~\ref{fig8}(b), whereas $\Omega_{yx}^{z}(\mathbf k)$ [Fig.~\ref{fig9}(d)] is dominated by a positive region around $\Gamma$, several sign-alternating rings, and broad negative weight at larger momenta. For $\mathbf m\parallel+z$, the response becomes more concentrated near $\Gamma$: $\Omega_{yx}^{y}(\mathbf k)$ [Fig.~\ref{fig9}(e)] consists mainly of irregular alternating rings with isolated boundary hot spots, while $\Omega_{yx}^{z}(\mathbf k)$ [Fig.~\ref{fig9}(f)] shows a broad positive center separated from the outer region by a negative annulus and additional alternating arcs. These changes in the distribution and cancellation of opposite SBC contributions account for the strong magnetization dependence of the integrated TR-even spin conductivity in Fig.~\ref{fig8}.

A distinct momentum-space character is found for the TR-odd intraband response in Figs.~\ref{fig9}(g)--\ref{fig9}(j). Unlike the broad SBC distributions, the odd response is concentrated mainly along narrow constant-energy contours, consistent with its Fermi-surface origin. For $\mathbf m\parallel+x$, $\sigma_{yx}^{y,\mathrm{odd}}(\mathbf k)$ in Fig.~\ref{fig9}(g) forms several intense, nearly continuous $\Gamma$-centered contours with alternating signs, together with additional arc-like features near the zone boundary. The $z$-polarized component in Fig.~\ref{fig9}(h) is weaker and more fragmented, appearing as disconnected positive and negative segments along several concentric contours. Both distributions are symmetric about $k_x=0$. For $\mathbf m\parallel+y$ [Fig.~\ref{fig9}(i)], $\sigma_{yx}^{x,\mathrm{odd}}(\mathbf k)$ forms a largely continuous contour pattern around $\Gamma$, with a pronounced negative inner region surrounded by positive contributions and additional positive weight near the Brillouin-zone boundary. For $\mathbf m\parallel+z$ [Fig.~\ref{fig9}(j)], the response also follows nested $\Gamma$-centered contours, but the positive and negative contributions are distributed over thinner and more weakly connected segments. The latter two distributions are symmetric about $k_y=0$, consistent with the $M_x\mathcal T$ constraint. At $E-E_F=+0.25$~eV (Fig.~\ref{fig15}), the TR-even spin Berry curvature retains broad sign-alternating distributions across the Brillouin zone, with pronounced concentric structures around $\Gamma$ whose detailed form depends on the magnetization direction and spin polarization. The TR-odd response, by contrast, remains confined to Fermi-surface-like contours, showing that the distinct momentum-space characters of the interband and intraband spin Hall responses persist away from the Fermi level.

\subsection{Magnetization-angle dependence of the spin Hall response}

The magnetization dependence of the transverse spin Hall response is further examined by continuously rotating the magnetization in the $xy$ and $xz$ planes. Figure~\ref{fig10} shows the angular evolution of $\sigma_{yx}^{x}$, $\sigma_{yx}^{y}$, and $\sigma_{yx}^{z}$ at the Fermi level for the $P_{\uparrow}$ state. The symmetry-adapted angular forms are derived in Appendix~F, and the corresponding fitting coefficients are summarized in Table~\ref{table3}.

For rotation in the $xy$ plane [Figs.~\ref{fig10}(a)--\ref{fig10}(f)], the three transverse components exhibit clearly different angular dependences. In the TR-even sector, $\sigma_{yx}^{x,\mathrm{even}}$ vanishes for $\mathbf m\parallel\pm x$ and $\mathbf m\parallel y$ but develops finite extrema at intermediate angles, whereas $\sigma_{yx}^{y,\mathrm{even}}$ varies between about $1$ and $9~(\hbar/e)(\mathrm{S/cm})$. The $z$-polarized component is much larger in magnitude and remains close to $-120~(\hbar/e)(\mathrm{S/cm})$ throughout the rotation. The TR-odd response displays a much stronger angular variation: $\sigma_{yx}^{x,\mathrm{odd}}$ reaches a magnitude exceeding $10^{3}~(\hbar/e)(\mathrm{S/cm})$ near $\phi=90^\circ$, while $\sigma_{yx}^{y,\mathrm{odd}}$ changes continuously from a large positive value at $\phi=0^\circ$ to a comparable negative value at $\phi=180^\circ$ and vanishes at $\phi=90^\circ$. The much smaller $\sigma_{yx}^{z,\mathrm{odd}}$ follows the same odd parity about $\phi=90^\circ$. These limits recover the symmetry constraints for $\mathbf m\parallel x$ and $\mathbf m\parallel y$ derived in Appendix~E. 

For the $xz$ rotation [Figs.~\ref{fig10}(g)--\ref{fig10}(l)], the angular anisotropy is again strongly component dependent. $\sigma_{yx}^{x,\mathrm{even}}$ changes sign across $\theta=90^\circ$ and reaches magnitudes of about $35~(\hbar/e)(\mathrm{S/cm})$ at intermediate angles, whereas $\sigma_{yx}^{z,\mathrm{even}}$ remains large and negative, varying between approximately $-120$ and $-160~(\hbar/e)(\mathrm{S/cm})$. The $\sigma_{yx}^{y,\mathrm{even}}$ component in Fig.~\ref{fig10}(h) varies much more weakly in magnitude but shows a noticeable non-sinusoidal profile. In contrast to most other components, which are already well reproduced by the low-order symmetry-allowed harmonics, this behavior suggests the presence of appreciable higher-order angular contributions beyond the lowest-order terms. In the odd sector, $\sigma_{yx}^{y,\mathrm{odd}}$ again dominates, changing from about $+1.5\times10^{3}$ to $-1.5\times10^{3}~(\hbar/e)(\mathrm{S/cm})$ between $\theta=0^\circ$ and $180^\circ$, while $\sigma_{yx}^{x,\mathrm{odd}}$ and $\sigma_{yx}^{z,\mathrm{odd}}$ remain much smaller. At $\theta=90^\circ$, the calculated components reduce to the $\mathbf m\parallel z$ selection rules, providing a direct connection between the continuous angular dependence and the high-symmetry results discussed above.

\begin{table}[htbp]
\centering
\caption{
Fitting coefficients describing the angular dependence of the transverse spin Hall conductivity for magnetization rotations in the $xy$ and $xz$ planes. The conductivities are given in units of $(\hbar/e)(\mathrm{S/cm})$.
}
\label{tab:SHC_angle_fit}
\begin{tabular}{cccc}
\hline\hline
Component & Coefficient & $xy$-plane rotation & $xz$-plane rotation \\
\hline

$\sigma_{yx}^{x,\mathrm{even}}$
& $A_{2}$ & $-5.7188$ & $-33.5518$ \\
& $A_{4}$ & $-2.0490$ & $-7.3828$ \\[2pt]

$\sigma_{yx}^{y,\mathrm{even}}$
& $B_{0}$ & $+3.4075$ & $+6.4657$ \\
& $B_{2}$ & $+2.7969$ & $+0.3189$ \\
& $B_{4}$ & $+2.5829$ & $+1.1744$ \\[2pt]

$\sigma_{yx}^{z,\mathrm{even}}$
& $C_{0}$ & $-117.7179$ & $-145.1469$ \\
& $C_{2}$ & $-2.0711$ & $+22.6763$ \\
& $C_{4}$ & $+0.8427$ & $+3.8929$ \\[2pt]

$\sigma_{yx}^{x,\mathrm{odd}}$
& $D_{1}$ & $-1244.9200$ & $-53.1815$ \\
& $D_{3}$ & $-6.9049$ & $+10.0311$ \\
& $D_{5}$ & $+10.5121$ & $-0.1741$ \\[2pt]

$\sigma_{yx}^{y,\mathrm{odd}}$
& $E_{1}$ & $+1413.1020$ & $+1416.2760$ \\
& $E_{3}$ & $+14.6247$ & $-0.8655$ \\
& $E_{5}$ & $-11.2625$ & $+0.4557$ \\[2pt]

$\sigma_{yx}^{z,\mathrm{odd}}$
& $F_{1}$ & $+55.3738$ & $+72.5129$ \\
& $F_{3}$ & $-2.3897$ & $-9.9990$ \\
& $F_{5}$ & $-2.2981$ & $-8.1075$ \\

\hline\hline
\end{tabular}
\label{table3}
\end{table}

\section{CONCLUSION} 

In summary, we have investigated the electrically and magnetically tunable charge--spin conversion in sliding-ferroelectric bilayer \ce{Fe3GaTe2}. The two degenerate polarization ground states originate from opposite interlayer sliding vectors and carry reversed out-of-plane dipole moments, with a calculated switching barrier of about $16.3$~meV per unit cell. The reduced $C_{3v}$ symmetry of the bilayer gives rise to a strongly magnetization-dependent Edelstein response. For an electric field along $+x$, $\delta S_{y,z}^{\mathrm{even/odd}}$ are allowed for $\mathbf m\parallel+x$, $\delta S_{y,z}^{\mathrm{even}}$ and $\delta S_x^{\mathrm{odd}}$ for $\mathbf m\parallel+y$, and only $\delta S_y^{\mathrm{even}}$ and $\delta S_x^{\mathrm{odd}}$ for $\mathbf m\parallel+z$. For fixed in-plane magnetization, the two ferroelectric states are related by $M_z\mathcal T$, leading to component-selective sign reversal: the in-plane TR-even components reverse sign while the corresponding TR-odd components are preserved, whereas the opposite relation applies to the $z$-polarized response. For $\mathbf m\parallel z$, the two states are related directly by $M_z$, so that the allowed $\delta S_y^{\mathrm{even}}$ and $\delta S_x^{\mathrm{odd}}$ both reverse sign. The response is also strongly energy dependent, with $\delta S_y^{\mathrm{even}}$ reaching nearly $2~\hbar$ for $\mathbf m\parallel+y$ and reversing sign within a narrow energy window on the electron-doping side. The element-resolved results identify Fe-$d$ states as the dominant source, with appreciable Te contributions. The momentum-resolved calculations provide a microscopic picture of these responses. The TR-even Edelstein contribution follows the constant-energy contours because of its predominantly intraband origin, whereas the TR-odd contribution is more localized in momentum space and reflects interband processes between nearby spin--orbit-coupled bands. Their distributions also retain the magnetic symmetry of each configuration: the allowed components are symmetric about $k_x=0$ for $\mathbf m\parallel+x$ and about $k_y=0$ for $\mathbf m\parallel+y$ and $+z$. Continuous magnetization rotations in the $xy$ and $xz$ planes connect these high-symmetry limits and reveal pronounced angular anisotropy together with symmetry-allowed higher harmonics.

 The spin-current response shows a parallel but distinct symmetry structure. For $\mathbf m\parallel+x$, both TR-even and TR-odd sectors allow $\sigma_{xx}^{x}$, $\sigma_{yx}^{y}$, and $\sigma_{yx}^{z}$. For $\mathbf m\parallel+y$ and $+z$, the TR-even sector retains the same three components, whereas the TR-odd sector allows $\sigma_{yx}^{x}$, $\sigma_{xx}^{y}$, and $\sigma_{xx}^{z}$; for $\mathbf m\parallel+z$, the additional $C_{3z}$ symmetry further imposes $\sigma_{yx}^{y,\mathrm{even}} =-\sigma_{xx}^{x,\mathrm{even}}$ and $\sigma_{yx}^{x,\mathrm{odd}} =\sigma_{xx}^{y,\mathrm{odd}}$. For fixed in-plane magnetization, the two ferroelectric states are related by $M_z\mathcal T$: the $x$- and $y$-polarized TR-even spin-conductivity components reverse sign while the corresponding TR-odd components remain unchanged, whereas the $z$-polarized components show the opposite behavior. For $\mathbf m\parallel z$, the two states are related directly by $M_z$, so that the $x$- and $y$-polarized components reverse sign in both time-reversal sectors, while the $z$-polarized components remain unchanged. The TR-even transverse spin Hall response reaches values of order $10^{3}(~\hbar/e)(\mathrm{S/cm})$ in selected energy regions, while the dominant TR-odd longitudinal response is considerably larger and is governed by intraband Fermi-surface transport. In momentum space, the TR-even interband response appears as a broad, sign-competing spin Berry curvature, whereas the TR-odd response follows narrow constant-energy contours. Both sectors obey the same mirror-axis change from $k_x=0$ for $\mathbf m\parallel+x$ to $k_y=0$ for $\mathbf m\parallel+y$ and $+z$.  
 
 These results establish sliding-ferroelectric bilayer \ce{Fe3GaTe2} as an effective multiferroic-metal platform for tunable charge--spin conversion. More broadly, they extend the functionality of metallic multiferroics to active control of nonequilibrium spin responses and provide a practical materials strategy for reconfiguring both spin accumulation and spin-current generation through ferroelectric switching and magnetization rotation.

\section*{ACKNOWLEDGMENTS}

This work was financially supported by  the  China Postdoctoral Science Foundation under Grant Number GZC20261825.

\begin{figure*}[t]
\includegraphics[width=170mm]{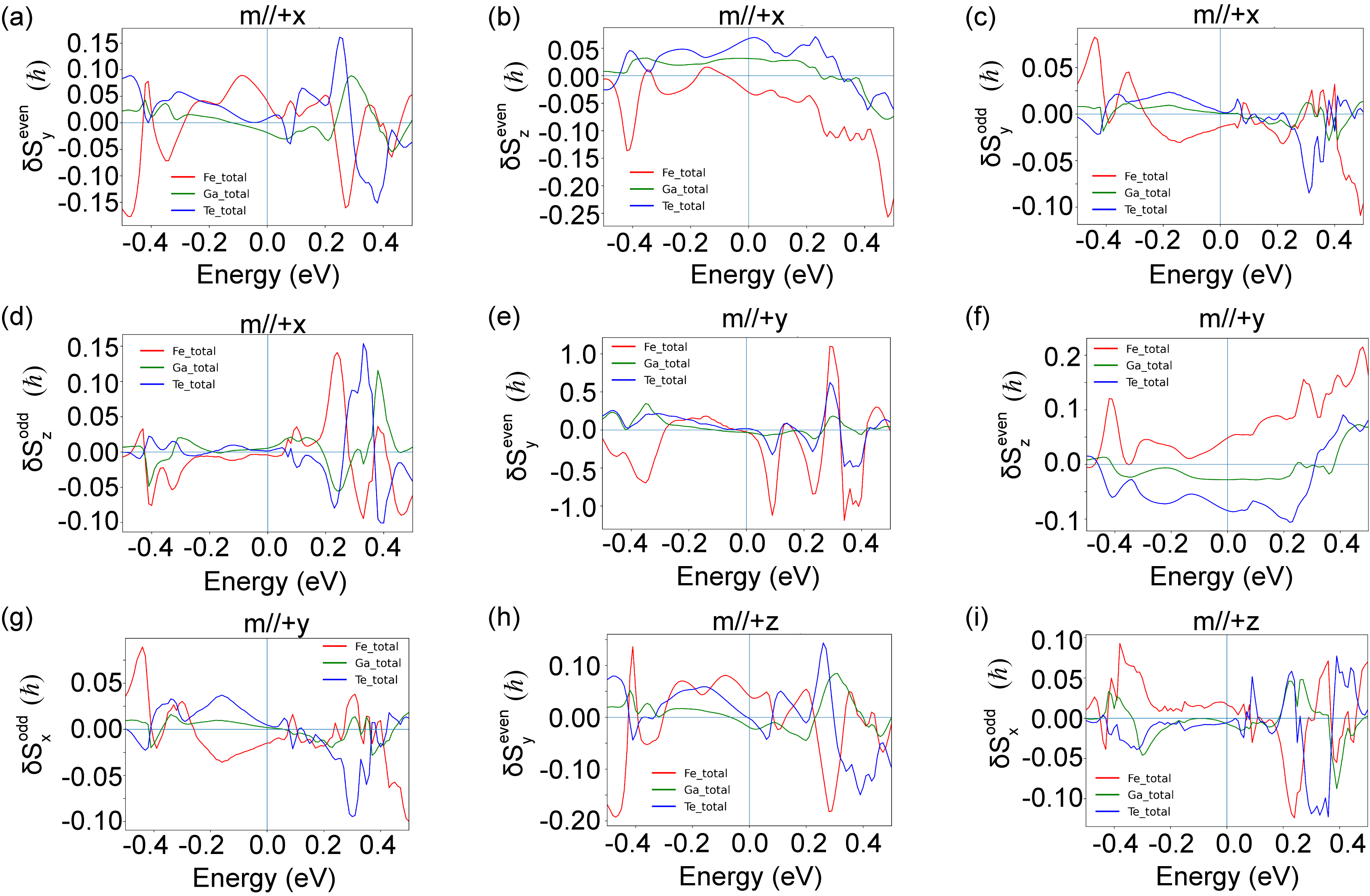}
\caption{ Element-resolved contributions to the current-induced spin accumulation in $P_{\uparrow}$ state under an electric field applied along the $+x$ direction, with the reference field $E_x=1$~V/\AA. (a)--(d) $\delta S_y^{\mathrm{even}}$, $\delta S_z^{\mathrm{even}}$, $\delta S_y^{\mathrm{odd}}$, and $\delta S_z^{\mathrm{odd}}$, respectively, for $\mathbf m\parallel +x$. (e)--(g) $\delta S_y^{\mathrm{even}}$, $\delta S_z^{\mathrm{even}}$, and $\delta S_x^{\mathrm{odd}}$, respectively, for $\mathbf m\parallel +y$. (h,i) $\delta S_y^{\mathrm{even}}$ and $\delta S_x^{\mathrm{odd}}$, respectively, for $\mathbf m\parallel +z$. Red, green, and blue curves denote the summed contributions from Fe, Ga, and Te atoms, respectively. }
\label{fig11}
\end{figure*}

\begin{figure*}[t]
\includegraphics[width=148mm]{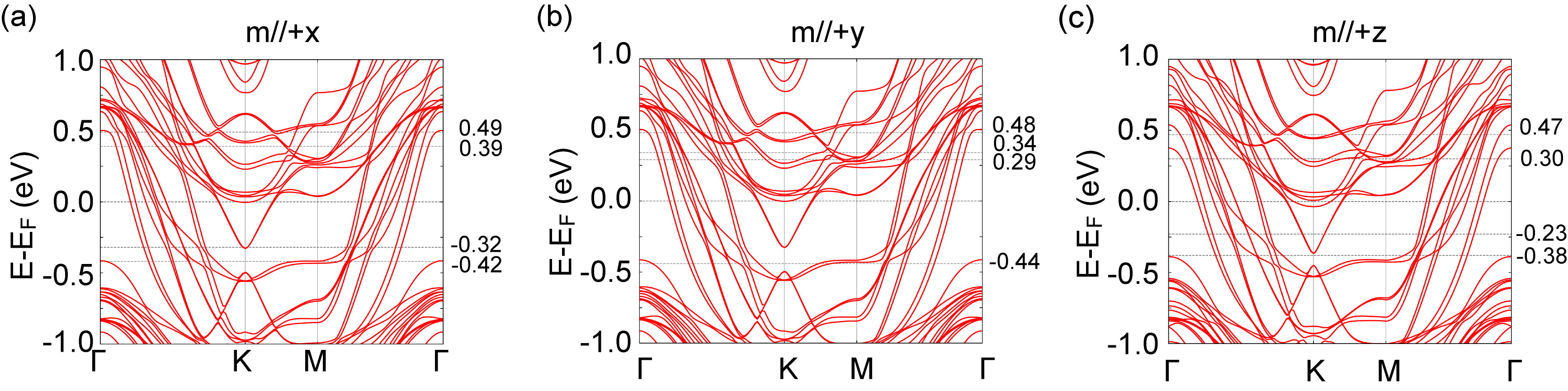}
\caption{ Electronic band structures of bilayer \ce{Fe3GaTe2} for (a) $\mathbf m\parallel +x$, (b) $\mathbf m\parallel +y$, and (c) $\mathbf m\parallel +z$ along the high-symmetry path $\Gamma$--K--M--$\Gamma$. The Fermi level is set to zero. The horizontal dashed lines indicate the representative energies selected from the energy-dependent Edelstein responses for the momentum-resolved analysis. For $\mathbf m\parallel +x$, the selected energies are $-0.42$, $-0.32$, $0.39$, and $0.49$~eV; for $\mathbf m\parallel +y$, they are $-0.44$, $0.29$, $0.34$, and $0.48$~eV; and for $\mathbf m\parallel +z$, they are $-0.38$, $-0.23$, $0.30$, and $0.47$~eV. }
\label{fig12}
\end{figure*}

\begin{figure}[t]
\includegraphics[width=85mm]{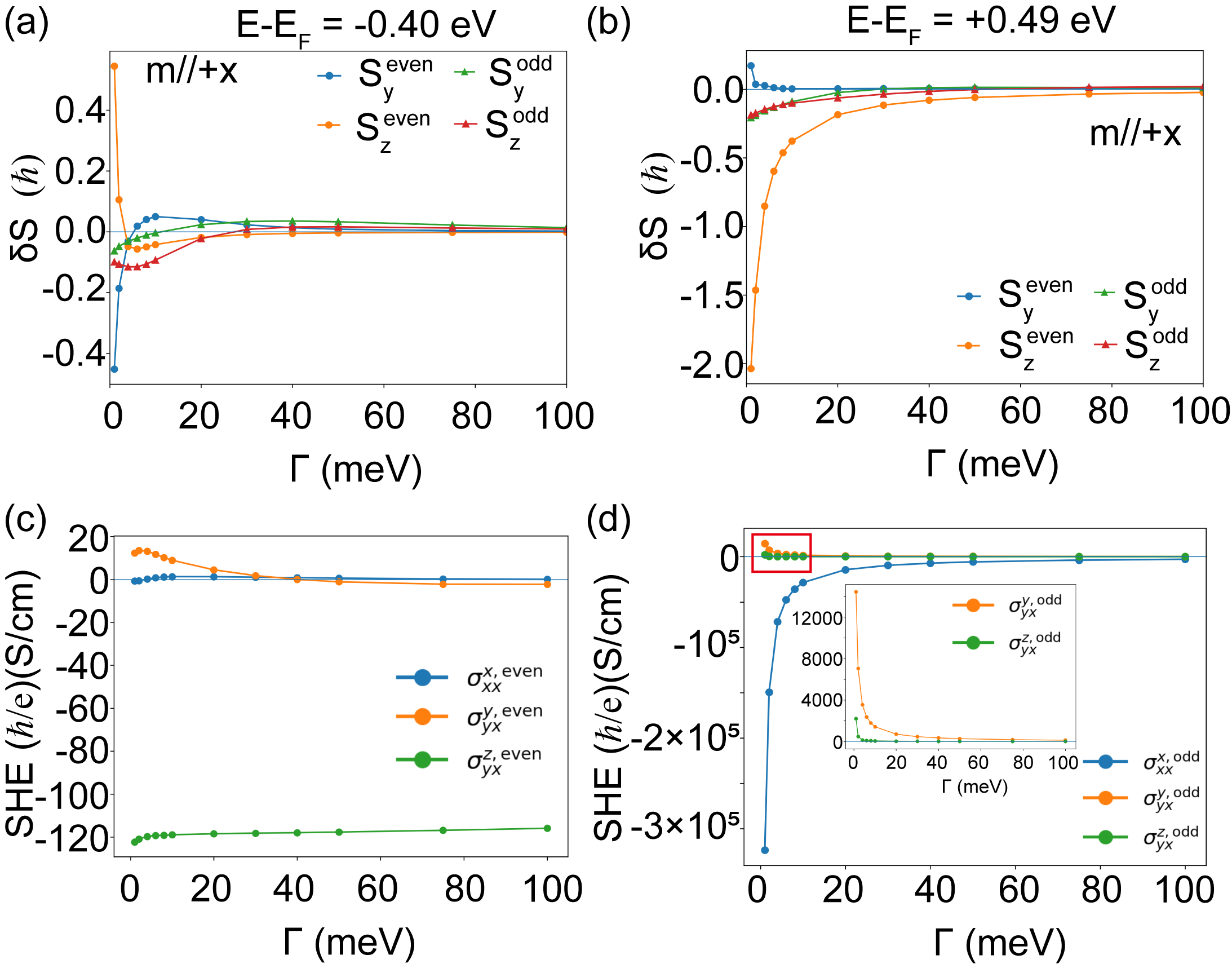}
\caption{ Dependence of the current-induced spin responses on the phenomenological broadening parameter $\Gamma$ in bilayer \ce{Fe3GaTe2} for $P_{\uparrow}$ state with $\mathbf m\parallel +x$. (a,b) $\Gamma$ dependence of $\delta S_y^{\mathrm{even}}$, $\delta S_z^{\mathrm{even}}$, $\delta S_y^{\mathrm{odd}}$, and $\delta S_z^{\mathrm{odd}}$ at $E-E_F=-0.40$ and $+0.49$~eV, respectively, under an electric field applied along $+x$, with the reference field $E_x=1$~V/\AA. (c,d) Corresponding $\Gamma$ dependence of the spin-conductivity components at the Fermi level. Panel (c) shows the TR-even interband response, whereas panel (d) shows the TR-odd intraband response. The inset in (d) enlarges the red-boxed region to resolve the comparatively smaller $\sigma_{yx}^{y,\mathrm{odd}}$ and $\sigma_{yx}^{z,\mathrm{odd}}$ components. The spin conductivities are given in units of $(\hbar/e)(\mathrm{S/cm})$ after applying the effective-thickness correction for the two-dimensional slab. }
\label{fig13}
\end{figure}

\begin{figure}[t]
\includegraphics[width=56mm]{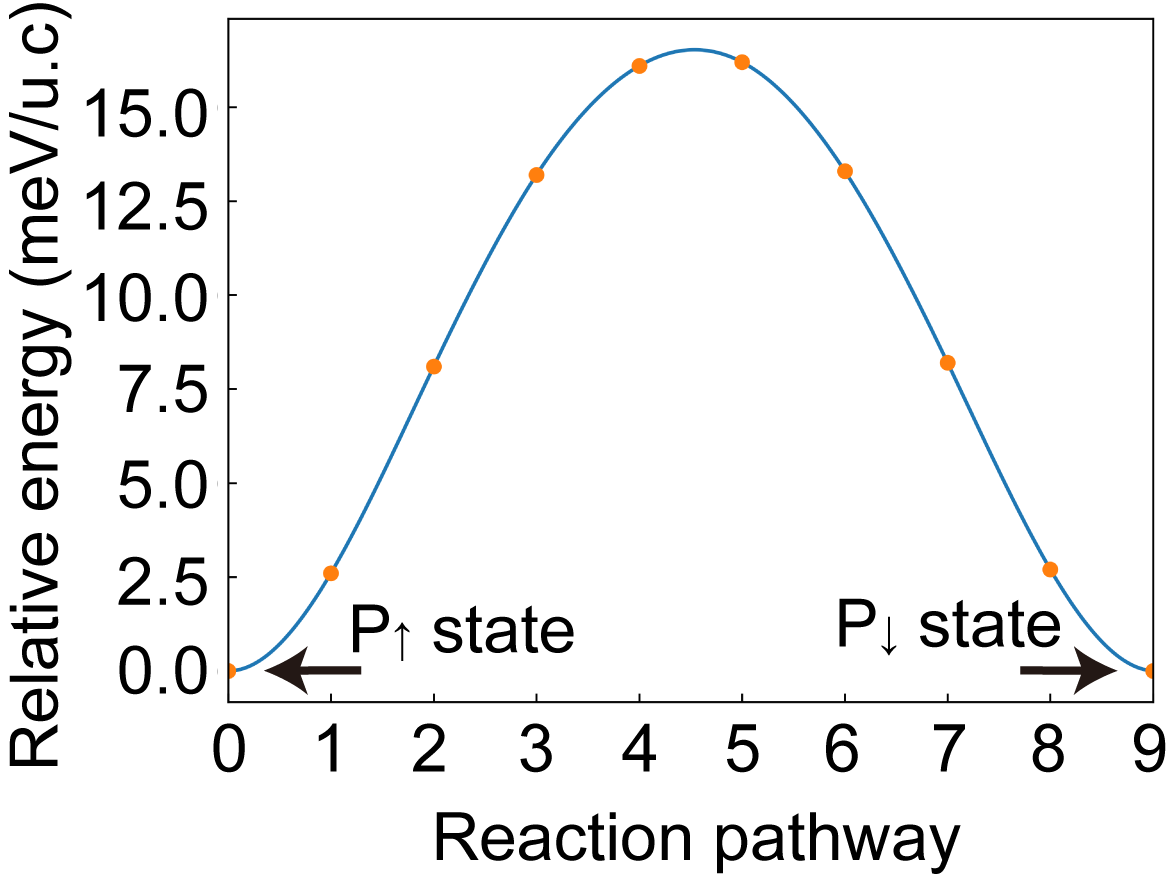}
\caption{ Calculated relative-energy profile along the interlayer sliding pathway between $P_{\uparrow}$ and $P_{\downarrow}$ states of bilayer \ce{Fe3GaTe2}. The two states are energetically degenerate and correspond to opposite ferroelectric polarizations.  }
\label{fig14}
\end{figure}

\begin{figure}[hbtp]
\includegraphics[width=85mm]{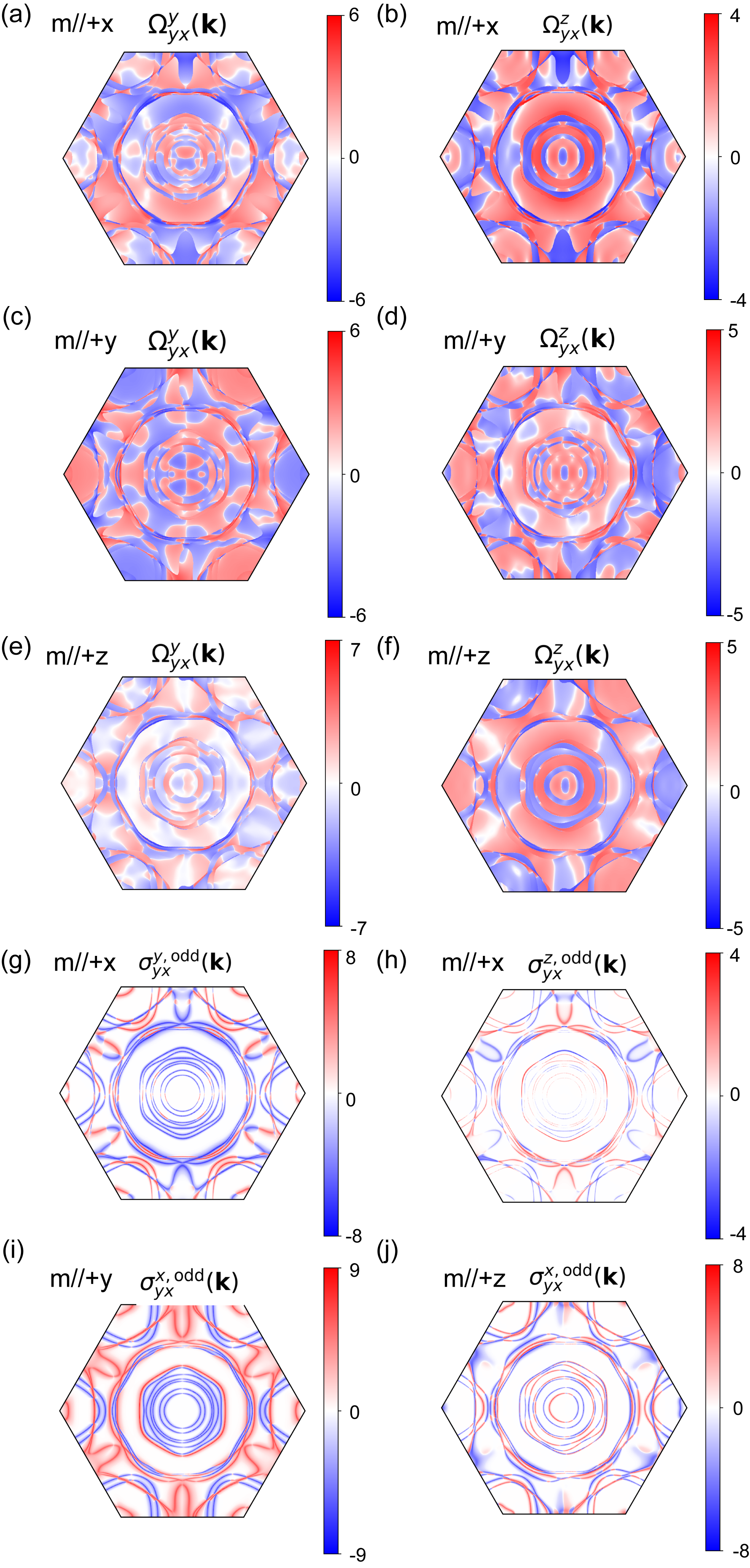}
\caption{ Momentum-resolved spin Hall response at $E-E_F=+0.25$ eV. Panels (a)--(f) show the spin Berry curvature associated with the TR-even interband intrinsic response. (a,b) $\Omega_{yx}^{y}(\mathbf k)$ and $\Omega_{yx}^{z}(\mathbf k)$ for $\mathbf m\parallel+x$; (c,d) corresponding results for $\mathbf m\parallel+y$; and (e,f) corresponding results for $\mathbf m\parallel+z$. Panels (g)--(j) show the momentum-resolved TR-odd intraband spin conductivity. (g,h) $\sigma_{yx}^{y,\mathrm{odd}}(\mathbf k)$ and $\sigma_{yx}^{z,\mathrm{odd}}(\mathbf k)$ for $\mathbf m\parallel+x$; (i) $\sigma_{yx}^{x,\mathrm{odd}}(\mathbf k)$ for $\mathbf m\parallel+y$; and (j) $\sigma_{yx}^{x,\mathrm{odd}}(\mathbf k)$ for $\mathbf m\parallel+z$. Red and blue indicate positive and negative contributions, respectively, while the hexagonal boundary marks the first Brillouin zone. The spin Berry curvature is given in units of \(\AA^2\). }
\label{fig15}
\end{figure}

\appendix

\section{Symmetry constraints on the Edelstein effect}

The bilayer \ce{Fe3GaTe2} crystallizes in the space group
$P3m1$ (No.~156), corresponding to the $C_{3v}$ point group.
Compared with the $D_{3h}$ symmetry of monolayer
\ce{Fe3GaTe2}~\cite{Brizolla2026SOT}, the horizontal mirror symmetry
is absent in the bilayer, thereby relaxing the symmetry constraints
on the Edelstein response.

Using the even/odd decomposition defined in Sec.~II, we consider the
response to an electric field applied along $+x$, so that only the
first column of the Edelstein tensor,
$(\chi_{xx},\chi_{yx},\chi_{zx})$, is relevant.
Under the vertical mirror
\[
M_x:(x,y,z)\rightarrow(-x,y,z),
\]
the electric field transforms as
\[
E_x\rightarrow-E_x,
\]
whereas the spin, being an axial vector, transforms as
\[
(S_x,S_y,S_z)\rightarrow(S_x,-S_y,-S_z).
\]
Therefore,
\[
(\chi_{xx},\chi_{yx},\chi_{zx})
\rightarrow
(-\chi_{xx},+\chi_{yx},+\chi_{zx})
\]
under the unitary $M_x$ operation.

For $\mathbf m\parallel+x$, the threefold rotational symmetry is broken
while $M_x$ remains a symmetry, and the magnetic point group reduces
to $m$. Consequently, $\chi_{xx}$ is forbidden in both the
time-reversal-even and time-reversal-odd sectors, whereas
$\chi_{yx}$ and $\chi_{zx}$ remain allowed. Thus, for an electric
field along $x$, the allowed responses are
$\delta S_y^{\mathrm{even}}$,
$\delta S_z^{\mathrm{even}}$,
$\delta S_y^{\mathrm{odd}}$, and
$\delta S_z^{\mathrm{odd}}$.

For $\mathbf m\parallel+y$, the mirror $M_x$ reverses the
magnetization and therefore survives only in the antiunitary form
$M_x\mathcal T$, giving the magnetic point group $m'$.
The TR-even sector retains the spatial transformation signs
$(-,+,+)$, whereas the TR-odd sector acquires an additional minus sign
and follows $(+,-,-)$. Hence, the allowed responses are
$\delta S_y^{\mathrm{even}}$,
$\delta S_z^{\mathrm{even}}$, and
$\delta S_x^{\mathrm{odd}}$.

For $\mathbf m\parallel+z$, the threefold rotation $C_{3z}$ remains a
unitary symmetry, whereas the vertical mirrors reverse the
out-of-plane magnetization and survive only when combined with time
reversal. The magnetic point group is therefore
$3m'=\langle C_{3z},M_x\mathcal T\rangle$.
In contrast to the in-plane magnetization cases, $C_{3z}$ mixes the
$x$ and $y$ components and must therefore be applied to the full
response tensor. Invariance under $C_{3z}$ requires
\begin{equation}
\chi=
\begin{pmatrix}
a & -b & 0\\
b & a & 0\\
0 & 0 & c
\end{pmatrix}.
\label{eq:EE_C3_general}
\end{equation}

The remaining $M_x\mathcal T$ symmetry imposes different constraints
on the two time-reversal sectors. In the even sector it requires
$a=c=0$, yielding
\begin{equation}
\chi^{\mathrm{even}}=
\begin{pmatrix}
0 & -b & 0\\
b & 0 & 0\\
0 & 0 & 0
\end{pmatrix},
\label{eq:EE_C3_even}
\end{equation}
whereas in the odd sector it requires $b=0$, giving
\begin{equation}
\chi^{\mathrm{odd}}=
\begin{pmatrix}
a & 0 & 0\\
0 & a & 0\\
0 & 0 & c
\end{pmatrix}.
\label{eq:EE_C3_odd}
\end{equation}
Therefore, for $E_x$, only
$\delta S_y^{\mathrm{even}}$ and
$\delta S_x^{\mathrm{odd}}$ remain finite.

The symmetry-allowed Edelstein responses for the three representative
magnetization directions are summarized in Table~IV.

\begin{table}[htbp]
\centering
\caption{
Symmetry constraints on the Edelstein response of the bilayer
with $C_{3v}$ symmetry for an electric field along $+x$.
For $\mathbf{m}\parallel+x$ and $+y$, the entries denote the
transformation signs of $(\chi_{xx},\chi_{yx},\chi_{zx})$.
}
\label{tab:bilayer_symmetry}
\renewcommand{\arraystretch}{1.15}
\setlength{\tabcolsep}{4pt}
\begin{tabular}{c c c c}
\hline\hline
$\mathbf{m}$ & Operation &
$\chi^{\mathrm{even}}_{ix}$ &
$\chi^{\mathrm{odd}}_{ix}$ \\
\hline

\multirow{2}{*}{$+x$}
& $M_x$
& $(-,+,+)$
& $(-,+,+)$ \\
& Allowed
& $\chi_{yx},\chi_{zx}$
& $\chi_{yx},\chi_{zx}$ \\
\hline

\multirow{2}{*}{$+y$}
& $M_x\mathcal{T}$
& $(-,+,+)$
& $(+,-,-)$ \\
& Allowed
& $\chi_{yx},\chi_{zx}$
& $\chi_{xx}$ \\
\hline

\multirow{3}{*}{$+z$}
& $C_{3z}$
& \multicolumn{2}{c}{
$\chi_{xx}=\chi_{yy},\
 \chi_{xy}=-\chi_{yx},\
 \chi_{zx}=0$
} \\
& $M_x\mathcal{T}$
& $a=c=0$
& $b=0$ \\
& Allowed
& $\chi_{yx}$
& $\chi_{xx}$ \\
\hline\hline
\end{tabular}
\end{table}

\section{Symmetry relation between two ferroelectric states}

The two ferroelectric configurations considered here, denoted as
$P_{\uparrow}$ and $P_{\downarrow}$, originate from opposite interlayer sliding vectors,
\[
\mathbf t_{P_{\uparrow}}=(1/3,-1/3),
\qquad
\mathbf t_{P_{\downarrow}}=(-1/3,1/3)=-\mathbf t_{P_{\uparrow}},
\]
and therefore carry opposite out-of-plane polarizations,
\[
P_{P_{\downarrow}}=-P_{P_{\uparrow}}.
\]

The two structures are related by the horizontal mirror operation
\[
M_z:(x,y,z)\rightarrow(x,y,-z),
\]
so that
\[
\mathrm{state\text{-}P_{\uparrow}}
\xleftrightarrow[]{M_z}
\mathrm{state\text{-}P_{\downarrow}}.
\]
Since both spin and magnetization are axial vectors,
\[
M_z:
(m_x,m_y,m_z)
\rightarrow
(-m_x,-m_y,m_z),
\]
and
\[
M_z:
(S_x,S_y,S_z)
\rightarrow
(-S_x,-S_y,S_z).
\]

For an out-of-plane magnetization,
$\mathbf m\parallel z$, $M_z$ preserves the magnetization direction
and directly relates the two ferroelectric states.
For an in-plane magnetization,
$\mathbf m\parallel x$ or $\mathbf m\parallel y$,
$M_z$ reverses the magnetization direction. Therefore, the two
ferroelectric states with the same in-plane magnetization are related
by the antiunitary operation $M_z\mathcal T$.

\subsection{Edelstein response}

For the electric-field geometry considered in this work,
$\mathbf E\parallel+x$, the horizontal mirror leaves the applied field
unchanged,
\[
M_z:E_x\rightarrow E_x.
\]
For fixed in-plane magnetization, the relation between $P_{\uparrow}$ and $P_{\downarrow}$ 
is therefore determined by $M_z\mathcal T$. Combining the spatial
transformation under $M_z$ with the even/odd behavior under
magnetization reversal gives
\begin{equation}
\begin{aligned}
\delta S_x^{\mathrm{even}}(P_{\downarrow})
&=-\delta S_x^{\mathrm{even}}(P_{\uparrow}),&
\delta S_x^{\mathrm{odd}}(P_{\downarrow})
&=\delta S_x^{\mathrm{odd}}(P_{\uparrow}),\\
\delta S_y^{\mathrm{even}}(P_{\downarrow})
&=-\delta S_y^{\mathrm{even}}(P_{\uparrow}),&
\delta S_y^{\mathrm{odd}}(P_{\downarrow})
&=\delta S_y^{\mathrm{odd}}(P_{\uparrow}),\\
\delta S_z^{\mathrm{even}}(P_{\downarrow})
&=\delta S_z^{\mathrm{even}}(P_{\uparrow}),&
\delta S_z^{\mathrm{odd}}(P_{\downarrow})
&=-\delta S_z^{\mathrm{odd}}(P_{\uparrow}).
\end{aligned}
\label{eq:FE_EE_inplane}
\end{equation}

For $\mathbf m\parallel z$, the two ferroelectric states with the same
magnetization direction are connected directly by $M_z$, without
time reversal. Hence, the even and odd sectors obey the same spatial
sign relation,
\begin{equation}
\begin{aligned}
\delta S_x^{\mathrm{even/odd}}(P_{\downarrow})
&=-\delta S_x^{\mathrm{even/odd}}(P_{\uparrow}),\\
\delta S_y^{\mathrm{even/odd}}(P_{\downarrow})
&=-\delta S_y^{\mathrm{even/odd}}(P_{\uparrow}),\\
\delta S_z^{\mathrm{even/odd}}(P_{\downarrow})
&=\delta S_z^{\mathrm{even/odd}}(P_{\uparrow}).
\end{aligned}
\label{eq:FE_EE_outofplane}
\end{equation}

Combining Eqs.~\eqref{eq:FE_EE_inplane} and
\eqref{eq:FE_EE_outofplane} with the symmetry-allowed components
derived in Appendix~A gives the ferroelectric-state relations shown
in Fig.~\ref{fig3}.

\subsection{Spin Hall response}

For the in-plane spin-current propagation considered here
($i=x,y$), the velocity components are unchanged under $M_z$,
\[
v_x\rightarrow v_x,
\qquad
v_y\rightarrow v_y.
\]
Using the spin-current operator defined in Eq.~(8), the transformation
is therefore determined by the spin-polarization direction,
\[
J_i^x\rightarrow-J_i^x,
\qquad
J_i^y\rightarrow-J_i^y,
\qquad
J_i^z\rightarrow+J_i^z.
\]
Since $E_x$ is unchanged, the corresponding spatial transformation is
\[
\sigma_{ix}^{x}\rightarrow-\sigma_{ix}^{x},
\qquad
\sigma_{ix}^{y}\rightarrow-\sigma_{ix}^{y},
\qquad
\sigma_{ix}^{z}\rightarrow+\sigma_{ix}^{z}.
\]

For fixed in-plane magnetization, $P_{\uparrow}$ and $P_{\downarrow}$ states are related by
$M_z\mathcal T$. The TR-even components therefore follow the spatial
signs above, whereas the TR-odd components acquire an additional minus
sign:
\begin{equation}
\begin{aligned}
\sigma_{ix}^{x,\mathrm{even}}(P_{\downarrow})
&=-\sigma_{ix}^{x,\mathrm{even}}(P_{\uparrow}),&
\sigma_{ix}^{x,\mathrm{odd}}(P_{\downarrow})
&=\sigma_{ix}^{x,\mathrm{odd}}(P_{\uparrow}),\\
\sigma_{ix}^{y,\mathrm{even}}(P_{\downarrow})
&=-\sigma_{ix}^{y,\mathrm{even}}(P_{\uparrow}),&
\sigma_{ix}^{y,\mathrm{odd}}(P_{\downarrow})
&=\sigma_{ix}^{y,\mathrm{odd}}(P_{\uparrow}),\\
\sigma_{ix}^{z,\mathrm{even}}(P_{\downarrow})
&=\sigma_{ix}^{z,\mathrm{even}}(P_{\uparrow}),&
\sigma_{ix}^{z,\mathrm{odd}}(P_{\downarrow})
&=-\sigma_{ix}^{z,\mathrm{odd}}(P_{\uparrow}).
\end{aligned}
\label{eq:FE_SHC_inplane}
\end{equation}

For $\mathbf m\parallel z$, the two states are connected directly by
$M_z$. Consequently, the even and odd sectors follow the same spatial
transformation,
\begin{equation}
\begin{aligned}
\sigma_{ix}^{x,\mathrm{even/odd}}(P_{\downarrow})
&=-\sigma_{ix}^{x,\mathrm{even/odd}}(P_{\uparrow}),\\
\sigma_{ix}^{y,\mathrm{even/odd}}(P_{\downarrow})
&=-\sigma_{ix}^{y,\mathrm{even/odd}}(P_{\uparrow}),\\
\sigma_{ix}^{z,\mathrm{even/odd}}(P_{\downarrow})
&=\sigma_{ix}^{z,\mathrm{even/odd}}(P_{\uparrow}),
\end{aligned}
\label{eq:FE_SHC_outofplane}
\end{equation}
where $i=x,y$.

\section{Angular dependence of the Edelstein response}
\label{app:angular_dependence}

To understand the angular dependence of the current-induced spin
accumulation, we follow the symmetry-adapted expansion of the Edelstein
response with respect to the magnetization direction. For the
time-reversal-even contribution,
$\delta S_i^{\mathrm{even}}=\sum_{j=x,y}
\chi_{ij}^{\mathrm{even}}E_j$, the response tensor is an even function
of the magnetization direction. Keeping terms up to fourth order gives

\begin{equation}
\begin{aligned}
\chi_{ij}^{\mathrm{even}}(\mathbf m)
&=
\chi_{ij}^{(0)}
\\
&\quad+
\sum_{ab}
\chi_{ij}^{(2),ab}m_am_b
\\
&\quad+
\sum_{abcd}
\chi_{ij}^{(4),abcd}m_am_bm_cm_d .
\end{aligned}
\label{eq:even_expansion}
\end{equation}

where $a,b,c,d=x,y,z$. Similarly, the time-reversal-odd response,
$\delta S_i^{\mathrm{odd}}=\sum_{j=x,y}
\chi_{ij}^{\mathrm{odd}}E_j$, is an odd function of the magnetization.
Up to fifth order,

\begin{equation}
\begin{aligned}
\chi_{ij}^{\mathrm{odd}}(\mathbf m)
&=
\sum_a
\chi_{ij}^{(1),a}m_a
+
\sum_{abc}
\chi_{ij}^{(3),abc}m_am_bm_c
\\
&\quad+
\sum_{abcde}
\chi_{ij}^{(5),abcde}m_am_bm_cm_dm_e .
\end{aligned}
\label{eq:odd_expansion}
\end{equation}

The separation into even and odd powers automatically incorporates
the behavior under magnetization reversal. The remaining coefficients
are constrained by the generators of the $C_{3v}$ point group,
namely the threefold rotation $C_{3z}$ and the vertical mirror $M_x$.
Their actions on the quantities entering the response are
\begin{equation}
\begin{aligned}
C_{3z}:\quad
\begin{pmatrix}
E_x\\ E_y
\end{pmatrix}
&\rightarrow
R_3
\begin{pmatrix}
E_x\\ E_y
\end{pmatrix},
\\
\begin{pmatrix}
m_x\\ m_y
\end{pmatrix}
&\rightarrow
R_3
\begin{pmatrix}
m_x\\ m_y
\end{pmatrix},
\qquad
m_z\rightarrow m_z,
\\
\begin{pmatrix}
\delta S_x\\ \delta S_y
\end{pmatrix}
&\rightarrow
R_3
\begin{pmatrix}
\delta S_x\\ \delta S_y
\end{pmatrix},
\qquad
\delta S_z\rightarrow\delta S_z,
\end{aligned}
\label{eq:C3_transform}
\end{equation}

where
\[
R_3=
\begin{pmatrix}
-\frac{1}{2} & -\frac{\sqrt{3}}{2}\\
\frac{\sqrt{3}}{2} & -\frac{1}{2}
\end{pmatrix}.
\]
and

\begin{equation}
\begin{aligned}
M_x:\quad
&E_x\rightarrow-E_x,\qquad E_y\rightarrow E_y,\\
&(m_x,m_y,m_z)\rightarrow(m_x,-m_y,-m_z),\\
&(\delta S_x,\delta S_y,\delta S_z)
\rightarrow
(\delta S_x,-\delta S_y,-\delta S_z).
\end{aligned}
\label{eq:Mx_transform}
\end{equation}

For every symmetry operation, Eqs.~(\ref{eq:even_expansion}) and
(\ref{eq:odd_expansion}) must retain the same form after simultaneously
transforming $\mathbf E$, $\mathbf m$, and $\delta\mathbf S$.
Comparing coefficients of equivalent powers of the magnetization removes
the symmetry-forbidden terms and relates coefficients that transform
into one another. Importantly, the $C_{3z}$ constraint must be imposed
before fixing $E_y=0$, because a pure $E_x$ field is rotated into a
linear combination of $E_x$ and $E_y$ under $C_{3z}$. After the symmetry
reduction, we set the applied electric field to
$\mathbf E=(E_x,0,0)$.

\subsection{Magnetization rotation in the $xy$ plane}

For magnetization rotating in the $xy$ plane, we write
\[
\mathbf m=(\cos\phi,\sin\phi,0),
\]
where $\phi=0$ and $\pi/2$ correspond to
$\mathbf m\parallel+x$ and $\mathbf m\parallel+y$, respectively.
Under $M_x$, $\phi\rightarrow-\phi$. Since $E_x$ changes sign,
$\delta S_x$ is odd in $\phi$, whereas
$\delta S_y$ and $\delta S_z$ are even.
Together with the even/odd magnetization expansions above, this
restricts the TR-even response to even angular harmonics and the
TR-odd response to odd angular harmonics.
After applying both $C_{3z}$ and $M_x$, the TR-even response through
fourth order is
\begin{equation}
\begin{aligned}
\delta S_x^{\mathrm{even}}
&=(A_2\sin2\phi+A_4\sin4\phi)E_x,\\
\delta S_y^{\mathrm{even}}
&=(B_0+B_2\cos2\phi+B_4\cos4\phi)E_x,\\
\delta S_z^{\mathrm{even}}
&=(C_2\cos2\phi+C_4\cos4\phi)E_x.
\end{aligned}
\label{eq:EE_xy_even}
\end{equation}
For the TR-odd response, retaining terms up to fifth order gives
\begin{equation}
\begin{aligned}
\delta S_x^{\mathrm{odd}}
&=(D_1\sin\phi+D_3\sin3\phi+D_5\sin5\phi)E_x,\\
\delta S_y^{\mathrm{odd}}
&=(D_1\cos\phi+E_3\cos3\phi-D_5\cos5\phi)E_x,\\
\delta S_z^{\mathrm{odd}}
&=(F_1\cos\phi+F_5\cos5\phi)E_x.
\end{aligned}
\label{eq:EE_xy_odd}
\end{equation}

The common coefficients $D_1$ and $D_5$ appearing in
$\delta S_x^{\mathrm{odd}}$ and
$\delta S_y^{\mathrm{odd}}$ are not independent fitting parameters.
Their relation, including the opposite sign of the fifth-order terms,
is imposed by $C_{3z}$. The same symmetry also removes the
$\cos3\phi$ contribution from $\delta S_z^{\mathrm{odd}}$.

\subsection{Magnetization rotation in the $xz$ plane}

For rotation in the $xz$ plane,
\[
\mathbf m=(\cos\theta,0,\sin\theta),
\]
where $\theta=0$ and $\pi/2$ correspond to
$\mathbf m\parallel+x$ and $\mathbf m\parallel+z$, respectively.
Although the $xz$ plane is not invariant under $C_{3z}$, the
threefold constraint must be imposed on the full response tensor
before setting $m_y=0$. Under $M_x$,
\[
(m_x,0,m_z)\rightarrow(m_x,0,-m_z),
\]
corresponding to $\theta\rightarrow-\theta$.
Thus, $\delta S_x$ is odd in $\theta$, whereas
$\delta S_y$ and $\delta S_z$ are even.
The TR-even response through fourth order is therefore
\begin{equation}
\begin{aligned}
\delta S_x^{\mathrm{even}}
&=(G_2\sin2\theta+G_4\sin4\theta)E_x,\\
\delta S_y^{\mathrm{even}}
&=(H_0+H_2\cos2\theta+H_4\cos4\theta)E_x,\\
\delta S_z^{\mathrm{even}}
&=(I_2\cos2\theta+I_4\cos4\theta)E_x,
\end{aligned}
\label{eq:EE_xz_even}
\end{equation}
whereas the TR-odd response is
\begin{equation}
\begin{aligned}
\delta S_x^{\mathrm{odd}}
&=(J_1\sin\theta+J_3\sin3\theta+J_5\sin5\theta)E_x,\\
\delta S_y^{\mathrm{odd}}
&=(K_1\cos\theta+K_3\cos3\theta+K_5\cos5\theta)E_x,\\
\delta S_z^{\mathrm{odd}}
&=(L_1\cos\theta+L_3\cos3\theta+L_5\cos5\theta)E_x.
\end{aligned}
\label{eq:EE_xz_odd}
\end{equation}

A common feature of the two rotation planes is the absence of a zeroth-order term in the full magnetization expansion of $\delta S_z^{\mathrm{even}}$. This follows from the threefold rotational constraint: an in-plane electric field cannot generate a magnetization-independent out-of-plane axial response in the $C_{3v}$ crystal. For the $xz$ rotation, this constraint further removes an independent pure $m_z^2$ contribution, so that $\delta S_z^{\mathrm{even}}$ vanishes at $\theta=\pi/2$, consistent with the $3m'$ symmetry for $\mathbf m\parallel z$.

At the high-symmetry angles, Eqs.~\eqref{eq:EE_xy_even}--
\eqref{eq:EE_xz_odd} reduce to the selection rules summarized in
Table~IV. The allowed harmonic content is therefore fixed by the
$C_{3v}$ symmetry and the time-reversal parity of the response,
whereas the numerical coefficients are determined by the microscopic
electronic structure.

\section{Effective-thickness correction for the spin Hall conductivity}
\label{app:SHC_thickness}

For a two-dimensional slab calculated using a periodic supercell,
the vacuum region artificially enlarges the cell volume and therefore
reduces the spin Hall conductivity when it is expressed in the
three-dimensional unit of
\((\hbar/e)(\mathrm{S/cm})\).
To remove this artificial dependence on the vacuum thickness, we
introduce an effective volume for the bilayer, following the treatment
commonly adopted for two-dimensional systems~\cite{Pan2026PtBi2}.

The supercell and effective volumes are written as
\[
V_{\rm cell}=A c,
\qquad
V_{\rm eff}=A d_{\rm eff},
\]
where \(A\) is the in-plane unit-cell area, \(c\) is the out-of-plane
lattice constant of the simulation cell, and \(d_{\rm eff}\) is the
effective thickness of the bilayer. Since the calculated spin Hall
conductivity is inversely proportional to the volume used in the Kubo
formula, the conductivity obtained using the full supercell volume is
rescaled according to
\begin{equation}
\sigma_{ij}^{k,\rm eff}
=
\sigma_{ij}^{k,\rm cell}
\frac{V_{\rm cell}}{V_{\rm eff}}
=
\sigma_{ij}^{k,\rm cell}
\frac{c}{d_{\rm eff}} .
\label{eq:SHC_thickness_correction}
\end{equation}

For the present bilayer, we take
\(d_{\rm eff}=12.96~\text{\AA}\), corresponding to the vertical
extent of the atomic slab, while the supercell height is
\(c=43.32~\text{\AA}\). The resulting correction factor is therefore
\[
\frac{c}{d_{\rm eff}}\simeq 3.34.
\]
All spin Hall conductivities reported in
\((\hbar/e)(\mathrm{S/cm})\) are corrected using
Eq.~\eqref{eq:SHC_thickness_correction}. This correction only changes the absolute conductivity scale and does
not affect its sign, energy dependence, symmetry relations, or the
relative momentum-space distribution of the spin Hall response.

\section{Symmetry constraints on the spin-conductivity tensor}

This analysis concerns the response tensor within a given
ferroelectric configuration and is therefore distinct from the
ferroelectric-domain transformation discussed in Appendix~B. Throughout this work, the electric field is applied along \(+x\).
We therefore set \(j=x\) and consider the components
\(\sigma_{ix}^{k}\). Since the present system is a two-dimensional
bilayer, we focus on the in-plane spin-current propagation directions
\(i=x,y\).

The polarization bilayer belongs to the \(C_{3v}\) point group, generated by
the threefold rotation \(C_{3z}\) and the vertical mirror \(M_x\).
Under \(M_x\),
\[
M_x:(x,y,z)\rightarrow(-x,y,z).
\]
The electric field and velocity are polar vectors and transform as
\[
E_x\rightarrow-E_x,
\qquad
(v_x,v_y,v_z)\rightarrow(-v_x,v_y,v_z),
\]
whereas spin is an axial vector,
\[
(S_x,S_y,S_z)
\rightarrow
(S_x,-S_y,-S_z).
\]

For a given SHC component \(\sigma_{ix}^{k}\), the spatial
transformation under \(M_x\) is therefore determined by
\begin{equation}
\sigma_{ix}^{k}
\rightarrow
\eta_k^{S}\eta_i^{v}\eta_x^{E}
\,\sigma_{ix}^{k},
\label{eq:Mx_SHC_sign}
\end{equation}
where
\[
\eta_x^{E}=-1,
\qquad
(\eta_x^{v},\eta_y^{v})=(-1,+1),
\]
and
\[
(\eta_x^{S},\eta_y^{S},\eta_z^{S})
=
(+1,-1,-1).
\]
For the six in-plane components relevant to \(E_x\), Eq.~\eqref{eq:Mx_SHC_sign}
gives
\[
\begin{array}{c|cccccc}
 & \sigma_{xx}^{x} & \sigma_{yx}^{x}
 & \sigma_{xx}^{y} & \sigma_{yx}^{y}
 & \sigma_{xx}^{z} & \sigma_{yx}^{z} \\ \hline
M_x
 & + & - & - & + & - & +
\end{array}.
\]
These signs form the basis for the magnetization-dependent selection
rules derived below.

\subsection{Magnetization along $+x$}

For $\mathbf m\parallel+x$, the mirror $M_x$ preserves the magnetic
configuration, and the magnetic point group reduces to
\[
m=\{E,M_x\}.
\]
Since $M_x$ is unitary, the even and odd sectors obey the same spatial
constraint. For $E_x$ and in-plane spin-current propagation, the
allowed components are
$\sigma_{xx}^{x}$,
$\sigma_{yx}^{y}$, and
$\sigma_{yx}^{z}$
in both sectors.

\subsection{Magnetization along $+y$}

For $\mathbf m\parallel+y$, the unitary mirror $M_x$ reverses the
magnetization, so that the magnetic configuration is preserved by the
antiunitary operation $M_x\mathcal T$,
\[
m'=\{E,M_x\mathcal T\}.
\]
The TR-even sector follows the same spatial constraint as under
$M_x$, whereas the TR-odd sector acquires an additional minus sign
under magnetization reversal. Consequently, the allowed even
components remain
$\sigma_{xx}^{x}$,
$\sigma_{yx}^{y}$, and
$\sigma_{yx}^{z}$,
whereas the allowed odd components are
$\sigma_{yx}^{x}$,
$\sigma_{xx}^{y}$, and
$\sigma_{xx}^{z}$.

\subsection{Magnetization along $+z$}

For $\mathbf m\parallel+z$, the threefold rotation $C_{3z}$ remains a
unitary symmetry, whereas the vertical mirror reverses the
out-of-plane magnetization and therefore survives only in the
antiunitary form $M_x\mathcal T$. The magnetic point group is
\[
3m'=\{ C_{3z},M_x\mathcal T\} .
\]

The $M_x\mathcal T$ constraint first gives the same sets of allowed
components as for $\mathbf m\parallel y$. The remaining $C_{3z}$
symmetry further relates tensor elements. Since $C_{3z}$ is a proper
rotation, polar and axial vectors transform with the same in-plane
rotation matrix,
\[
R_3=
\begin{pmatrix}
-\frac{1}{2} & -\frac{\sqrt{3}}{2}\\
\frac{\sqrt{3}}{2} & -\frac{1}{2}
\end{pmatrix}.
\]
The spin-conductivity tensor therefore satisfies
\begin{equation}
\sigma_{ij}^{k}
=
(R_3)_{ii'}
(R_3)_{jj'}
(R_3)_{kk'}
\sigma_{i'j'}^{k'} .
\label{eq:SHC_C3}
\end{equation}
This constraint must be imposed on the full in-plane response tensor
before setting $E_y=0$, because $C_{3z}$ mixes the $x$ and $y$
components of both the electric field and the spin current.

Combining $C_{3z}$ with $M_x\mathcal T$ gives, in the TR-even sector,
\begin{equation}
\sigma_{yx}^{y,\mathrm{even}}
=
-\sigma_{xx}^{x,\mathrm{even}},
\label{eq:SHC_C3_even}
\end{equation}
while in the TR-odd sector,
\begin{equation}
\sigma_{yx}^{x,\mathrm{odd}}
=
\sigma_{xx}^{y,\mathrm{odd}}.
\label{eq:SHC_C3_odd}
\end{equation}
The remaining symmetry-allowed components
$\sigma_{yx}^{z,\mathrm{even}}$ and
$\sigma_{xx}^{z,\mathrm{odd}}$
are independent of these relations.

The complete set of symmetry-allowed in-plane spin-conductivity
components for $E\parallel x$ is therefore
\begin{equation}
\begin{aligned}
\mathbf m\parallel x:\quad&
\begin{cases}
\mathrm{even}:&
\sigma_{xx}^{x},\,
\sigma_{yx}^{y},\,
\sigma_{yx}^{z},\\
\mathrm{odd}:&
\sigma_{xx}^{x},\,
\sigma_{yx}^{y},\,
\sigma_{yx}^{z},
\end{cases}
\\[3pt]
\mathbf m\parallel y:\quad&
\begin{cases}
\mathrm{even}:&
\sigma_{xx}^{x},\,
\sigma_{yx}^{y},\,
\sigma_{yx}^{z},\\
\mathrm{odd}:&
\sigma_{yx}^{x},\,
\sigma_{xx}^{y},\,
\sigma_{xx}^{z},
\end{cases}
\\[3pt]
\mathbf m\parallel z:\quad&
\begin{cases}
\mathrm{even}:&
\sigma_{xx}^{x},\,
\sigma_{yx}^{y},\,
\sigma_{yx}^{z},
\quad
\sigma_{yx}^{y}=-\sigma_{xx}^{x},\\
\mathrm{odd}:&
\sigma_{yx}^{x},\,
\sigma_{xx}^{y},\,
\sigma_{xx}^{z},
\quad
\sigma_{yx}^{x}=\sigma_{xx}^{y}.
\end{cases}
\end{aligned}
\label{eq:SHC_summary}
\end{equation}
These relations provide the symmetry basis for selecting the
spin-conductivity components analyzed below.

\section{Angular dependence of the spin Hall response}
\label{app:SHC_angle}

Using the symmetry constraints derived in Appendix~E, we now consider
the angular dependence of the transverse spin-conductivity components
$\sigma_{yx}^{\alpha}$ ($\alpha=x,y,z$) for continuous magnetization
rotations in the $xy$ and $xz$ planes.

We write
\[
\mathbf m=(\cos\phi,\sin\phi,0)
\]
for rotation in the $xy$ plane and
\[
\mathbf m=(\cos\theta,0,\sin\theta)
\]
for rotation in the $xz$ plane. For convenience, we introduce
\[
\lambda=
\begin{cases}
\phi, & xy\text{-plane rotation},\\
\theta, & xz\text{-plane rotation}.
\end{cases}
\]
Under the vertical mirror $M_x$, both rotation paths satisfy
$\lambda\rightarrow-\lambda$. The transverse spin-conductivity
components transform as
\[
\left(
\sigma_{yx}^{x},
\sigma_{yx}^{y},
\sigma_{yx}^{z}
\right)
\rightarrow
\left(
-\sigma_{yx}^{x},
+\sigma_{yx}^{y},
+\sigma_{yx}^{z}
\right).
\]
Thus, $\sigma_{yx}^{x}$ is odd in $\lambda$, whereas
$\sigma_{yx}^{y}$ and $\sigma_{yx}^{z}$ are even.
Magnetization reversal corresponds to
$\lambda\rightarrow\lambda+\pi$; the TR-even response is invariant
under this operation, while the TR-odd response changes sign.
These constraints determine the allowed angular harmonics.

The TR-even transverse components through fourth order are
\begin{equation}
\begin{aligned}
\sigma_{yx}^{x,\mathrm{even}}(\lambda)
&=
A_2\sin2\lambda
+
A_4\sin4\lambda,\\
\sigma_{yx}^{y,\mathrm{even}}(\lambda)
&=
B_0
+
B_2\cos2\lambda
+
B_4\cos4\lambda,\\
\sigma_{yx}^{z,\mathrm{even}}(\lambda)
&=
C_0
+
C_2\cos2\lambda
+
C_4\cos4\lambda,
\end{aligned}
\label{eq:SHC_angle_even}
\end{equation}
whereas the TR-odd components, retaining harmonics up to fifth order,
are
\begin{equation}
\begin{aligned}
\sigma_{yx}^{x,\mathrm{odd}}(\lambda)
&=
D_1\sin\lambda
+
D_3\sin3\lambda
+
D_5\sin5\lambda,\\
\sigma_{yx}^{y,\mathrm{odd}}(\lambda)
&=
E_1\cos\lambda
+
E_3\cos3\lambda
+
E_5\cos5\lambda,\\
\sigma_{yx}^{z,\mathrm{odd}}(\lambda)
&=
F_1\cos\lambda
+
F_3\cos3\lambda
+
F_5\cos5\lambda.
\end{aligned}
\label{eq:SHC_angle_odd}
\end{equation}

At $\lambda=0$ and $\pi/2$,
Eqs.~\eqref{eq:SHC_angle_even} and~\eqref{eq:SHC_angle_odd}
reduce to the high-symmetry selection rules summarized in
Eq.~\eqref{eq:SHC_summary}.
The $C_{3z}$ constraint acts on the complete rank-three
spin-conductivity tensor, as discussed in Appendix~E, and does not
further reduce the harmonic forms above for the three fixed transverse
components considered here. The fitting coefficients are determined
independently for the $xy$ and $xz$ rotation planes and are summarized
in Table~III.

\section{Momentum-space symmetry of the Edelstein and spin Hall responses}
\label{app:momentum_symmetry}

\subsection{Momentum-resolved Edelstein response}

The magnetic symmetry constrains not only the Brillouin-zone-integrated
Edelstein response derived in Appendix~A, but also its distribution in
momentum space. For the electric-field geometry considered throughout
this work, the momentum-resolved spin accumulation is written as
\begin{equation}
\delta S_i(\mathbf{k})
=
\chi_{ix}(\mathbf{k})E_x,
\qquad
\delta S_i
=
\frac{1}{N_k}
\sum_{\mathbf{k}}
\delta S_i(\mathbf{k}),
\label{eq:kresolved_EE}
\end{equation}
where the response is decomposed into time-reversal-even and
time-reversal-odd contributions.

For $\mathbf m\parallel+x$, the unitary mirror $M_x$ acts in momentum
space as
\[
M_x:(k_x,k_y)\rightarrow(-k_x,k_y).
\]
Using the transformation signs derived in Appendix~A, the
symmetry-allowed components satisfy
\begin{equation}
\begin{aligned}
\chi_{yx}^{\mathrm{even/odd}}(-k_x,k_y)
&=
\chi_{yx}^{\mathrm{even/odd}}(k_x,k_y),\\
\chi_{zx}^{\mathrm{even/odd}}(-k_x,k_y)
&=
\chi_{zx}^{\mathrm{even/odd}}(k_x,k_y).
\end{aligned}
\label{eq:kEE_mx}
\end{equation}
Thus,
$\delta S_y^{\mathrm{even/odd}}(\mathbf{k})$ and
$\delta S_z^{\mathrm{even/odd}}(\mathbf{k})$ are symmetric with
respect to the $k_x=0$ line. By contrast, the symmetry-forbidden
$x$ component is antisymmetric under $k_x\rightarrow-k_x$ and
therefore vanishes after Brillouin-zone integration.

For $\mathbf m\parallel+y$, the relevant magnetic symmetry is the
antiunitary operation $M_x\mathcal T$. Since time reversal reverses
the crystal momentum, its combined action is
\[
M_x\mathcal T:
(k_x,k_y)\rightarrow(k_x,-k_y).
\]
The symmetry-allowed components therefore obey
\begin{equation}
\begin{aligned}
\chi_{yx}^{\mathrm{even}}(k_x,-k_y)
&=
\chi_{yx}^{\mathrm{even}}(k_x,k_y),\\
\chi_{zx}^{\mathrm{even}}(k_x,-k_y)
&=
\chi_{zx}^{\mathrm{even}}(k_x,k_y),\\
\chi_{xx}^{\mathrm{odd}}(k_x,-k_y)
&=
\chi_{xx}^{\mathrm{odd}}(k_x,k_y).
\end{aligned}
\label{eq:kEE_my}
\end{equation}
Hence, the corresponding momentum-resolved spin accumulations are
symmetric with respect to the $k_y=0$ line.

For $\mathbf m\parallel+z$, the magnetic point group is
$3m'=\langle C_{3z},M_x\mathcal T\rangle$.
The $M_x\mathcal T$ operation imposes the same momentum mapping,
yielding
\begin{equation}
\begin{aligned}
\delta S_y^{\mathrm{even}}(k_x,-k_y)
&=
\delta S_y^{\mathrm{even}}(k_x,k_y),\\
\delta S_x^{\mathrm{odd}}(k_x,-k_y)
&=
\delta S_x^{\mathrm{odd}}(k_x,k_y).
\end{aligned}
\label{eq:kEE_mz}
\end{equation}
The mirror axis of the momentum-resolved Edelstein response therefore
changes from $k_x=0$ for $\mathbf m\parallel x$ to $k_y=0$ for
$\mathbf m\parallel y,z$. The remaining $C_{3z}$ symmetry acts on the complete
momentum-resolved Edelstein tensor and mixes different tensor
components. An individual component is therefore not required to
display an independent threefold rotational pattern.

A response component forbidden after Brillouin-zone integration need
not vanish locally at each $\mathbf{k}$ point. Instead, magnetic
symmetry may impose an antisymmetric momentum-space distribution whose
contributions cancel between symmetry-related points.

\subsection{Momentum-space symmetry of the spin Hall response}

Having established the symmetry-allowed spin-conductivity components
in Appendix~E, we further examine their momentum-space distributions.
For the intrinsic interband spin Hall response, the conductivity can
be expressed in terms of the spin Berry curvature (SBC) as
\begin{equation}
\sigma_{yx}^{\alpha}
=
\frac{e}{2VN_k}
\sum_{\mathbf{k}}
\Omega_{yx}^{\alpha}(\mathbf{k}),
\label{eq:SBC_sum}
\end{equation}
where $\alpha=x,y,z$ denotes the spin-polarization direction,
$V$ is the effective volume, and $N_k$ is the number of sampled
$\mathbf{k}$ points. The momentum-resolved spin Berry curvature is
defined as
\begin{equation}
\begin{aligned}
\Omega_{yx}^{\alpha}(\mathbf k)
={}&
\sum_n f_{n\mathbf k}~\hbar^2
\sum_{m\neq n}
\\
&\quad
\frac{
-2\,{\rm Im}
\left[
\left\langle n\mathbf k
\left|
\frac{1}{2}
\left\{
\hat{\sigma}_{\alpha},\hat v_y
\right\}
\right|
m\mathbf k
\right\rangle
\left\langle
m\mathbf k
\left|
\hat v_x
\right|
n\mathbf k
\right\rangle
\right]
}{
\left(
\varepsilon_{n\mathbf k}
-
\varepsilon_{m\mathbf k}
\right)^2
}.
\end{aligned}
\label{eq:SBC_definition}
\end{equation}
Here $f_{n\mathbf{k}}$ is the Fermi--Dirac distribution,
$\hat{\sigma}_{\alpha}$ is the Pauli matrix for spin component
$\alpha$, and $\hat v_x$ and $\hat v_y$ are the velocity operators.

The spin Berry curvature describes the intrinsic interband contribution to the spin conductivity. In the small-broadening limit, it takes the conventional Berry-curvature form with the $(\varepsilon_{n\mathbf{k}}-\varepsilon_{m\mathbf{k}})^{-2}$ energy denominator and is determined by virtual transitions between different Bloch bands. The TR-odd contribution, by contrast, arises from intraband Fermi-surface transport. Accordingly, the TR-even response in Fig.~\ref{fig9} is represented by $\Omega_{yx}^{y}(\mathbf{k})$ and $\Omega_{yx}^{z}(\mathbf{k})$, corresponding to $\sigma_{yx}^{y,\mathrm{even}}$ and $\sigma_{yx}^{z,\mathrm{even}}$, respectively. The TR-odd response is shown directly as the momentum-resolved spin conductivity, with $\sigma_{yx}^{y,\mathrm{odd}}(\mathbf{k})$ and $\sigma_{yx}^{z,\mathrm{odd}}(\mathbf{k})$ for $\mathbf m\parallel x$, and $\sigma_{yx}^{x,\mathrm{odd}}(\mathbf{k})$ for $\mathbf m\parallel y,z$. Only the transverse components with spin-current flow along $y$ are considered here. For $\mathbf m\parallel x$, symmetry allows $\sigma_{yx}^{y,\mathrm{odd}}$ and $\sigma_{yx}^{z,\mathrm{odd}}$, whereas for $\mathbf m\parallel y,z$ only $\sigma_{yx}^{x,\mathrm{odd}}$ remains in the transverse geometry. Other symmetry-allowed odd components correspond to longitudinal spin-current flow along $x$ and are not included in the present spin Hall analysis.

The momentum-space distributions inherit the tensor transformation
properties derived in Appendix~E, together with the corresponding
mapping of the crystal momentum. For $\mathbf m\parallel+x$, the
unitary mirror $M_x$ gives
\[
(k_x,k_y)\rightarrow(-k_x,k_y),
\]
and the two TR-even SBC components satisfy
\begin{equation}
\Omega_{yx}^{y,z}(-k_x,k_y)
=
\Omega_{yx}^{y,z}(k_x,k_y).
\label{eq:SBC_mx}
\end{equation}
Thus, both distributions are symmetric with respect to the $k_x=0$
line.

For $\mathbf m\parallel y$ and $\mathbf m\parallel z$, the
antiunitary symmetry $M_x\mathcal T$ gives
\[
(k_x,k_y)\rightarrow(k_x,-k_y),
\]
and therefore
\begin{equation}
\Omega_{yx}^{y,z}(k_x,-k_y)
=
\Omega_{yx}^{y,z}(k_x,k_y).
\label{eq:SBC_myz}
\end{equation}
The mirror axis of the TR-even momentum-space distribution therefore
changes from $k_x=0$ for $\mathbf m\parallel x$ to $k_y=0$ for
$\mathbf m\parallel y,z$.

The TR-odd intraband spin conductivity obeys the corresponding
momentum-space constraints. For $\mathbf m\parallel x$,
\begin{equation}
\begin{aligned}
\sigma_{yx}^{y,\mathrm{odd}}(-k_x,k_y)
&=
\sigma_{yx}^{y,\mathrm{odd}}(k_x,k_y),\\
\sigma_{yx}^{z,\mathrm{odd}}(-k_x,k_y)
&=
\sigma_{yx}^{z,\mathrm{odd}}(k_x,k_y),
\end{aligned}
\label{eq:odd_SHC_mx}
\end{equation}
whereas for $\mathbf m\parallel y$ and
$\mathbf m\parallel z$,
\begin{equation}
\sigma_{yx}^{x,\mathrm{odd}}(k_x,-k_y)
=
\sigma_{yx}^{x,\mathrm{odd}}(k_x,k_y).
\label{eq:odd_SHC_myz}
\end{equation}
Hence, the TR-odd momentum-resolved spin conductivity has the same
mirror axis as the corresponding TR-even intrinsic response:
$k_x=0$ for $\mathbf m\parallel x$ and $k_y=0$ for
$\mathbf m\parallel y,z$.

For $\mathbf m\parallel z$, the additional $C_{3z}$ symmetry acts on
the complete rank-three spin-conductivity tensor and simultaneously
mixes the spin-polarization, spin-current, and electric-field indices.
Consequently, an individual component such as
$\Omega_{yx}^{y}(\mathbf{k})$ or
$\Omega_{yx}^{z}(\mathbf{k})$ is not required to exhibit an
independent threefold rotational pattern.

\section{Momentum-space representation}
\label{app:SBC_representation}

The momentum-resolved spin Hall response was evaluated on a dense
two-dimensional \(k\)-point mesh. We focus on
\(\Omega_{yx}^{y}(\mathbf{k})\) and
\(\Omega_{yx}^{z}(\mathbf{k})\), corresponding to the two transverse
spin Hall components considered in this work.

The calculated data are initially defined in the primitive reciprocal
cell spanned by the in-plane reciprocal lattice vectors
\(\mathbf b_1\) and \(\mathbf b_2\),
\[
\mathbf k=u\mathbf b_1+v\mathbf b_2,
\qquad
0\leq u,v<1.
\]
For visualization, the response is mapped into the hexagonal first
Brillouin zone using reciprocal-lattice periodicity,
\[
\Omega_{yx}^{\alpha}(\mathbf k+\mathbf G)
=
\Omega_{yx}^{\alpha}(\mathbf k),
\qquad
\mathbf G=n_1\mathbf b_1+n_2\mathbf b_2.
\]
Thus, points outside the chosen first-Brillouin-zone representation are
translated by reciprocal lattice vectors to their equivalent positions.
This procedure only changes the representation of the momentum-space
distribution and does not modify the calculated response.

To display both weak and strongly enhanced contributions on the same
scale, we use the signed-log representation
\begin{equation}
\widetilde{\Omega}
=
\begin{cases}
{\rm sgn}(\Omega)\log_{10}|\Omega|,
& |\Omega|>10,\\[4pt]
\Omega/10,
& |\Omega|\leq10.
\end{cases}
\label{eq:SBC_signed_log}
\end{equation}
Positive and negative contributions are represented by red and blue,
respectively.

\section{Switching the polarization states}
\label{app:broadening}
The CI-NEB minimum-energy path between $P_{\uparrow}$ and $P_{\downarrow}$ is shown in Fig.~\ref{fig14}. A formal coherent-switching estimate gives $E_c=\Delta E_b/|p_z|$, where $\Delta E_b$ is the barrier per unit cell and $p_z$ is the out-of-plane dipole moment \cite{Pan2026PtBi2}. Using $\Delta E_b\simeq16.3$~meV/u.c. and $|p_z|=2.19\times10^{-4}\,e\AA$ yields $E_c\simeq74$~V/\AA. This value should be regarded as an upper-bound estimate for homogeneous switching rather than a device coercive field, since ferroelectric reversal generally proceeds through spatially nonuniform processes such as domain nucleation and domain-wall propagation \cite{Shin2007DomainWall}. In van der Waals ferroelectrics, atomistic studies further indicate that local dipole-switching events can provide additional pathways that avoid a fully coherent reversal \cite{He2023CIPS}. Such local switching mechanisms can substantially reduce the field required to reverse the macroscopic polarization. A similar distinction between the intrinsic coherent-switching estimate and experimentally accessible switching conditions has been discussed for other two-dimensional ferroelectrics. For monolayer \ce{CuCrSe2}, first-principles calculations predict a sizable switching barrier and polarization \cite{Zhong2020NSR}, while related estimates for two-dimensional ferroelectrics suggest that the intrinsic switching field can reach the order of $10^{2}$~V/nm \cite{Pan2026PtBi2}. Nevertheless, room-temperature PFM measurements have demonstrated polarization reversal in single-layer \ce{CuCrSe2}, with a coercive bias of approximately 4--5~V \cite{Sun2024CuCrSe2}. Ferroelectric switching has likewise been experimentally demonstrated in metallic few-layer $1T'$-\ce{WTe2} under electrostatic gating \cite{Fei2018WTe2}. These observations indicate that the large coherent-switching field estimated for bilayer \ce{Fe3GaTe2} does not necessarily preclude experimentally accessible polarization reversal, although the actual switching field will depend on the microscopic nucleation pathway, domain structure, and device environment.
%\clearpage

\section{Dependence of interlayer magnetic coupling on the van der Waals Gap}

To further examine the robustness of the interlayer magnetic coupling in bilayer \ce{Fe3GaTe2}, we investigated its dependence on the van der Waals (vdW) gap. Since the interlayer separation can vary upon structural relaxation and may be affected by changes in the interlayer interaction, the magnetic ground state was evaluated for a series of vdW gaps around the optimized value. As summarized in Table~\ref{tab:FGaT_vdw_gap}, the fully relaxed bilayer has a vdW gap of approximately 3.095~\AA. Over the investigated range of 2.995--3.195~\AA, the FM configuration remains energetically favored over the AFM configuration, with $E_{\mathrm{AFM}}-E_{\mathrm{FM}}$ remaining positive throughout. The energy difference gradually decreases from 5.037 to 2.545~meV per unit cell as the vdW gap increases, indicating a progressive weakening of the interlayer magnetic coupling with increasing layer separation. Nevertheless, no reversal of the magnetic ground state occurs within this range, demonstrating that the FM interlayer coupling is robust against moderate variations of the vdW gap.

\begin{table}[htbp]
\centering
\caption{Total energies of bilayer \ce{Fe3GaTe2} for different vdW gaps, comparing the interlayer ferromagnetic (FM) and antiferromagnetic (AFM) configurations for in-plane magnetization. In the FM configuration, the magnetic moments of the two layers are aligned parallel, whereas in the AFM configuration they are aligned antiparallel. $\Delta E=E_{\mathrm{AFM}}-E_{\mathrm{FM}}$, with a positive value indicating that the interlayer FM configuration is energetically favored. The optimized vdW gap is highlighted in bold.}
\label{tab:FGaT_vdw_gap}
\begin{tabular}{cccc}
\hline\hline
vdW gap (\AA) 
& $E_{\mathrm{FM}}$ (eV) 
& $E_{\mathrm{AFM}}$ (eV) 
& $\Delta E$ (meV/cell) \\
\hline
2.995 & $-79.933$ & $-79.928$ & 5.04 \\
3.045 & $-79.935$ & $-79.931$ & 4.30 \\
\textbf{3.095} & $\mathbf{-79.936}$ & $\mathbf{-79.932}$ & $\mathbf{3.64}$ \\
3.145 & $-79.935$ & $-79.932$ & 3.05 \\
3.195 & $-79.933$ & $-79.930$ & 2.55 \\
\hline\hline
\end{tabular}
\end{table}

\section{Dependence on the broadening parameter}

The linear-response calculations employ a phenomenological broadening of $\Gamma=10$~meV. Within the convention $\tau=~\hbar/(2\Gamma)$, this value corresponds to a characteristic relaxation time of approximately $33$~fs, which lies in the femtosecond timescale relevant to charge and spin transport in two-dimensional systems~\cite{Abdukayumov2025PRL,Kempt2024PtSe2}. A broadening of $10$~meV has also been adopted in recent first-principles calculations of current-induced spin responses in two-dimensional systems~\cite{Pan2026VTe2,Pan2026PtBi2}.

To examine the influence of this parameter, we first calculate representative Edelstein spin accumulations as a function of $\Gamma$ for the $P_{\uparrow}$ ferroelectric state with $\mathbf m\parallel+x$ at $E-E_F=-0.40$ and $+0.49$~eV. As shown in Figs.~\ref{fig13}(a) and \ref{fig13}(b), the TR-even Edelstein response increases strongly as $\Gamma$ is reduced, reflecting the approximately $1/\Gamma$ dependence of the intraband Fermi-surface contribution in the small-broadening regime~\cite{Pan2026VTe2}. By contrast, the TR-odd response, which originates from interband processes, shows a much weaker broadening dependence and approaches the intrinsic limit as $\Gamma$ decreases.

We further examine the broadening dependence of the spin conductivity at the intrinsic Fermi level for the same $\mathbf m\parallel+x$ configuration. In this case, the microscopic assignment of the two time-reversal sectors is reversed: the TR-even contribution is interband in origin, whereas the TR-odd contribution is dominated by intraband Fermi-surface transport. As shown in Fig.~\ref{fig13}(c), the dominant TR-even transverse component $\sigma_{yx}^{z,\mathrm{even}}$ depends only weakly on $\Gamma$, varying from approximately $-122$ to $-116~(~\hbar/e)(\mathrm{S/cm})$ over the range $\Gamma=1$--$100$~meV. The smaller $\sigma_{yx}^{y,\mathrm{even}}$ component exhibits a stronger variation because of the partial cancellation among positive and negative interband contributions, while $\sigma_{xx}^{x,\mathrm{even}}$ remains small throughout.

A qualitatively different behavior is found for the TR-odd spin conductivity [Fig.~\ref{fig13}(d)]. The dominant longitudinal component $\sigma_{xx}^{x,\mathrm{odd}}$ and the transverse components $\sigma_{yx}^{y,\mathrm{odd}}$ and $\sigma_{yx}^{z,\mathrm{odd}}$ increase rapidly as $\Gamma$ is reduced. Over the converged broadening range, their dependence is well described by the intraband scaling \begin{equation} \sigma_{ij}^{\alpha,\mathrm{odd}} \propto \frac{1}{\Gamma}, \end{equation} consistent with the Fermi-surface form discussed in Sec.~III~E. For example, $\sigma_{xx}^{x,\mathrm{odd}}$ changes from approximately $-1.44\times10^{4}$ at $\Gamma=20$~meV to $-2.86\times10^{4}$ at $10$~meV and $-7.18\times10^{4}~(~\hbar/e)(\mathrm{S/cm})$ at $4$~meV. Deviations appearing at the smallest broadenings, particularly near $\Gamma=1$--$2$~meV for the weaker transverse components, may reflect the increasingly stringent Brillouin-zone sampling required to resolve the narrow Fermi-surface features.

The broadening tests therefore confirm the expected distinction between intraband and interband responses. The value $\Gamma=10$~meV used throughout the main calculations lies in a regime where the TR-odd spin conductivity follows the expected intraband scaling, while the dominant intrinsic TR-even spin Hall component is already only weakly dependent on $\Gamma$. It therefore provides a consistent reference for comparing the different ferroelectric states and magnetization configurations.

\section{Layer-resolved Edelstein response}

\begin{figure*}[hbtp]
\includegraphics[width=170mm]{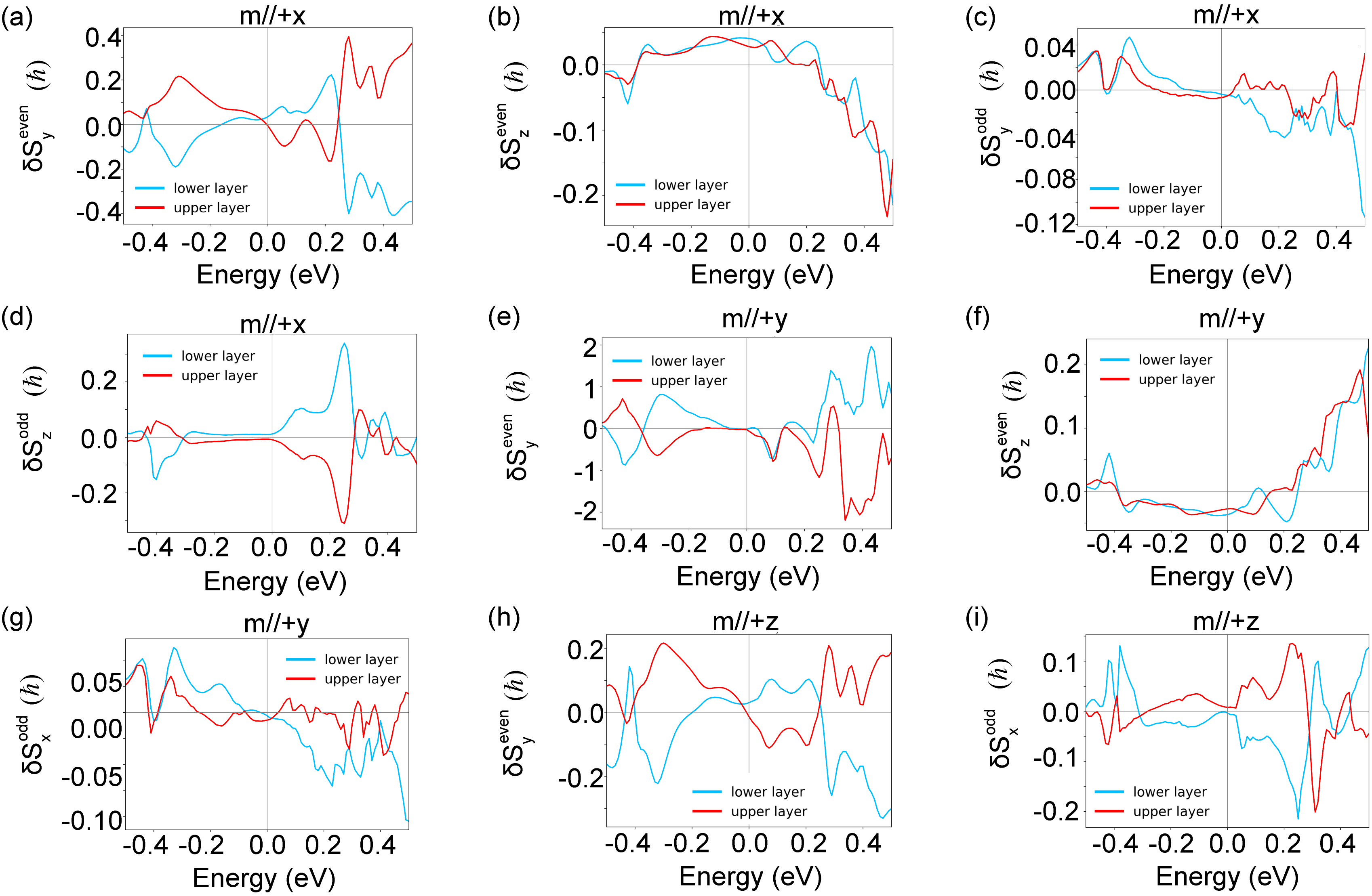}
\caption{ Layer-resolved current-induced spin accumulation in bilayer \ce{Fe3GaTe2} for the $P_{\uparrow}$ polarization state under an electric field applied along $+x$, with the reference field $E_x=1$~V/\AA. (a--d) Layer-resolved $\delta S_y^{\mathrm{even}}$, $\delta S_z^{\mathrm{even}}$, $\delta S_y^{\mathrm{odd}}$, and $\delta S_z^{\mathrm{odd}}$, respectively, for $\mathbf m\parallel+x$. (e--g) Corresponding $\delta S_y^{\mathrm{even}}$, $\delta S_z^{\mathrm{even}}$, and $\delta S_x^{\mathrm{odd}}$ for $\mathbf m\parallel+y$. (h,i) Corresponding $\delta S_y^{\mathrm{even}}$ and $\delta S_x^{\mathrm{odd}}$ for $\mathbf m\parallel+z$. Blue and red curves denote the lower- and upper-layer contributions, respectively. The total spin accumulation shown in Fig.~\ref{fig2} is obtained by summing the contributions from the two layers. }
\label{fig16}
\end{figure*}

Figure~\ref{fig16} shows the layer-resolved current-induced spin accumulation for the same magnetization configurations considered in Fig.~\ref{fig2}. The total Edelstein response can be decomposed as \[ \delta S_i^{\mathrm{tot}} = \delta S_i^{\mathrm{lower}} + \delta S_i^{\mathrm{upper}}. \] Because the two layers are inequivalent in the sliding-ferroelectric bilayer, their individual responses generally differ in both magnitude and sign.

For $\mathbf m\parallel+x$, the most pronounced layer contrast occurs in $\delta S_y^{\mathrm{even}}$ [Fig.~\ref{fig16}(a)]. The upper- and lower-layer contributions have opposite signs over broad energy ranges, particularly on the electron-doping side, so that the total response results from substantial interlayer cancellation. By contrast, $\delta S_z^{\mathrm{even}}$ [Fig.~\ref{fig16}(b)] shows similar energy dependence and mostly the same sign on the two layers, leading to a more constructive contribution to the total response. The TR-odd components also exhibit appreciable layer asymmetry, with $\delta S_z^{\mathrm{odd}}$ showing particularly strong opposite-sign features around $E-E_F\sim0.2$--$0.3$~eV [Fig.~\ref{fig16}(d)].

For $\mathbf m\parallel+y$, the layer contrast becomes especially strong for $\delta S_y^{\mathrm{even}}$ [Fig.~\ref{fig16}(e)]. On the electron-doping side, the two layers develop large responses of opposite sign, reaching magnitudes of order $1$--$2~\hbar$ in selected energy regions. The pronounced total response in Fig.~\ref{fig2}(c) therefore reflects the incomplete cancellation between two substantially larger layer-resolved contributions. In comparison, $\delta S_z^{\mathrm{even}}$ [Fig.~\ref{fig16}(f)] shows a more similar response on the two layers.

A comparable layer asymmetry is also present for $\mathbf m\parallel+z$. Both $\delta S_y^{\mathrm{even}}$ and $\delta S_x^{\mathrm{odd}}$ can contain sizable opposite-sign contributions from the two layers [Figs.~\ref{fig16}(h) and \ref{fig16}(i)]. These results show that the total Edelstein response can conceal substantial layer-localized spin accumulation. Consequently, surface-sensitive probes or asymmetric device geometries may detect a response that differs substantially from the layer-summed value shown in Fig.~\ref{fig2}.

\bibliography{bilayer}

\end{document}